\documentclass[preprint]{elsarticle}

\usepackage{graphicx}
\usepackage[dvipsnames,table]{xcolor}

\usepackage{blindtext}
\usepackage{tcolorbox}

\usepackage{wrapfig}
\usepackage{lipsum}

\usepackage{longtable}

\usepackage{lscape}

\usepackage[normalem]{ulem}

\usepackage{multirow}

\usepackage{array}
\newcolumntype{H}{>{\setbox0=\hbox\bgroup}c<{\egroup}@{}}

\usepackage{enumerate}

\usepackage{subfig}

\usepackage{array}
\newcolumntype{x}[1]{>{\centering\arraybackslash}p{#1}}

\usepackage{enumitem}

\usepackage{adjustbox}

\usepackage{hhline}

\usepackage{url}

\usepackage{array}
\usepackage{makecell}

\usepackage[multiple]{footmisc}

\usepackage{framed}

\newcommand{\fsr}{FSR\xspace}
\newcommand{\fsrs}{FSRs\xspace}

\newcommand{\cs}{case study\xspace}

\newcommand{\approach}{method\xspace}
\newcommand{\approaches}{
methods\xspace}
\newcommand{\Approach}{
Method\xspace}

\newcommand{\an}{a\xspace}

\newcommand{\ve}{\color{black} }
\newcommand{\pl}{\cellcolor[rgb]{.22,.33,.57} \color{white}}

\usepackage{tcolorbox}
\newtcolorbox{mybox}[2]{
    arc=0pt,
    boxrule=1pt,
    colback=#1,
    width=\linewidth,
}

\newcommand{\skr}[1]{}
\newcommand{\rv}[1]{ }

\begin{document}
\title{A Functional Safety Assessment Method for Cooperative Automotive Architecture\tnoteref{t1}}
\tnotetext[t1]{This work is a part of the i-CAVE research programme (14897 P14-18) funded by NWO
(Netherlands Organisation for Scientific Research).}

\author{Sangeeth Kochanthara\corref{cor1}}
\ead{s.kochanthara@tue.nl}

\author{Niels Rood}
\ead{n.rood@tue.nl}

\author{Arash Khabbaz Saberi}
\ead{a.khabbaz.saberi@tue.nl}

\author{Loek Cleophas}
\ead{l.g.w.a.cleophas@tue.nl}

\author{Yanja Dajsuren}
\ead{y.dajsuren@tue.nl}

\author{Mark van den Brand}
\ead{m.g.j.v.d.brand@tue.nl}

 \cortext[cor1]{Corresponding author}
 
 \affiliation{organization={Eindhoven University of Technology}, 
                 country={The Netherlands}}

\begin{abstract}

The scope of automotive functions
has grown from a single vehicle as an entity to multiple vehicles working together as an entity, referred to as cooperative driving. 
The current automotive safety standard, ISO 26262, is designed for single vehicles. With the increasing number of  cooperative driving capable vehicles on the road, it is now imperative to systematically 
assess the functional safety of architectures of these vehicles.   
Many methods are proposed to assess architectures with respect to different quality attributes in the software architecture domain, but to the best of our knowledge, functional safety assessment of automotive architectures is not explored in the literature.  
We present \an \approach, that leverages existing research in software architecture and safety engineering domains, to check whether the functional safety requirements for a cooperative driving scenario are fulfilled in the technical architecture of a 
vehicle. We  apply our \approach on a real-life academic prototype for a cooperative driving scenario, platooning, and discuss our insights.
\end{abstract}


\maketitle

\section{Introduction}
\label{sec:introduction}

Traffic congestion was estimated to cost 305 billion dollars in 2017 to traffic participants in the United States of America.\footnote{\scriptsize\url{https://www.smartcitiesdive.com/news/gridlock-woes-traffic-congestion-by-the-numbers/519959/}}
With continuously increasing urban population~\cite{alvarez2017considering}, traffic congestion will continue to be an inevitable problem for the foreseeable future. Around 70\% of all goods transported around the United States are moved by trucks, and the lion's share of the cost for operating trucks comprises fuel costs and driver salary~\cite{trego2010analysis}.
One potential solution to reduce traffic congestion and such operational costs is \textit{cooperative driving}.

\emph{Cooperative driving} refers to the collective optimization of the  traffic participants' behavior by sharing information 
using wireless communication such as a peer-to-peer network or via other actors like the cloud~\cite{ploeg2014analysis}.
Cooperative driving 
can improve traffic efficiency, reduces cost, and increases comfort~\cite{davila2013report,liang2015heavy,pelliccione2020beyond}. 
It is one of the 54 trends shaping the technology market, according to  market research.\footnote{\scriptsize\url{https://go.abiresearch.com/lp-54-technology-trends-to-watch-in-2020}}
In the year 2020 alone, 10.46 million new vehicles, with some form of cooperative driving capabilities, are projected to hit the roads.\footnote{\scriptsize\url{https://bit.ly/volkswagen-includes-nxp-v2x}}
With millions of cooperative driving capable vehicles on roads, the safety of these vehicles needs urgent attention.\footnote{\scriptsize\url{https://www.sciencedaily.com/releases/2019/05/190519191641.htm}}

A majority of the cooperative driving functionalities are achieved by determining a vehicle's behavior for optimal traffic behavior according to the information received from other traffic participants. 
Such optimal behaviors are achieved (partially or fully) using software-controlled steering, acceleration, and braking~\cite{dajsuren2019safety}.
Therefore, any problem in the software can lead to catastrophic effects not only to the vehicle itself but also to other traffic participants. 
To avoid such events, cooperative driving systems  are designed to operate in case of  failure or fail safely.

The current guidelines to ensure the safety of automotive systems (and their architecture) are provided by ISO 26262:2018 - a product development standard for the automotive domain~\cite{ISO26262-2}.
The ISO 26262 standard offers  systematic methods from the safety engineering domain  to identify safety requirements. 
Any automotive software architecture that fulfills these safety requirements is deemed safe-by-design.

ISO 26262 standard neither considers cooperative driving nor prescribes methods for architecture assessment.
The standard is designed for single vehicles and does not include a cooperative perspective in which a set of vehicles is seen as a single entity~\cite{mallozzi2019autonomous,nilsson2013functional}. 
This can mean that a low-risk safety requirement from a single-vehicle perspective can have catastrophic effects on other cooperating vehicles~\cite{pelliccione2020beyond}.
To create a functionally safe architecture from a cooperative perspective, existing studies have extended the standard guidelines~\cite{kochanthara2020semi,saberi2018functional} or presented an architecture framework~\cite{pelliccione2020beyond}. Yet, checking the safety of  software architecture of an existing vehicle for cooperative driving, remains an open question.

ISO 26262 standard does not prescribe methods to assess the  \textit{functional safety} of  automotive architecture. 
Many  approaches to assess architectures with respect to quality attributes have emerged in the software architecture domain in the past three decades~\cite{babar2004framework,dobrica2002survey,kazman1998architecture,bass2012software,bengtsson1998scenario,stoermer2003scam,bergner2005dosam,harrison2010pattern}.  
However, only some of these methods are designed for operational quality attributes like performance (in contrast to development quality attributes like maintainability)~\cite{bosch1999software,babar2004framework}. To the best of our knowledge,  none of these methods are  designed to assess the operational quality attribute \textit{functional safety} of automotive systems.

This paper presents \an \approach to assess the functional safety of 
 existing automotive architecture   for cooperative driving, by combining methods from the safety engineering and software architecture domains.
Our \approach has two parts: \\
$(i)$ derive Functional Safety Requirements (FSRs) for cooperative driving scena\-rios
---an extension of our earlier work~\cite{kochanthara2020semi};\\
$(ii)$ check whether the (technical) software architecture fulfills the derived functional safety requirements---a combination of techniques~\cite{wu2004safety,preschern2013building,kazman1998architecture} adapted from the software architecture domain.

\skr{The scope of our work is limited to the concept phase of ISO 26262. The technical safety requirement is not discussed in the concept phase, rather it is discussed in the product development phase and thus is beyond the scope of our work. We have revised the introduction section to define the scope of our work.}

This paper primarily focuses on the design phase (concept development phase in ISO 26262) and validation of the resultant requirements in the software architecture in the final product. 
We apply our \approach on the architecture of an academic prototype capable of cooperative driving. 
The cooperative driving scenario used for demonstration of our \approach is \emph{platooning}, in which 
a manually driven vehicle is autonomously followed by a train of vehicles.


\skr{We added a paragraph at the end of the Introduction section on the organization of the paper.

P2, please consider adding a description of the paper structure at the end of the introduction, i.e., "The rest of this paper is organized as follows ..."
}

The rest of the paper is organized as follows. 
Section~\ref{sec:background} presents the background relevant to the study. Section~\ref{sec:methodology} describes the proposed method to derive FSRs and check for their fulfillment in vehicles' technical software architecture. 
Section~\ref{sec:case_study} details the application of the proposed approach on an academic prototype for the cooperative driving use case, platooning, and interpreting the results from this case study. 
Section~\ref{sec:discussion} discusses our implicit assumptions, applicability, and scope of our approach.
Section~\ref{sec:related_work} outlines related research. 
Section~\ref{sec:threats_to_validity} presents threats to validity, followed by future research directions in Section~\ref{sec:future_work} and the conclusion in  Section~\ref{sec:conclusion}.

\section{Background} 
\label{sec:background}

In this section, we discuss the three basic concepts upon which we build the contributions of this paper. 
First, we outline  the relevant concepts in automotive functional safety. 
Second, we discuss some  basics  on safety tactics and patterns. 
Last, we give a brief introduction to the two views of automotive architecture.

\subsection{Functional safety}
Functional safety is defined as \emph{``an absence of unreasonable risk due to hazards caused by malfunctioning behaviour  of E/E systems"}~\cite{ISO26262-2} where E/E systems refer to electrical and/or electronic systems.
In the automotive domain, functional safety is defined by two standards: ISO 26262:2018 and ISO 21448~\cite{ISO21448}, serving complementary purposes. The former focuses on the hazards caused by the malfunctioning of components of a system, while the latter does on the hazards resulting from the functional insufficiency and misuse~\cite{ISO26262-2,ISO21448}. 
ISO 26262~\cite{ISO26262-2} is the current safety standard with its latest revision from 2018. In contrast, ISO 21448~\cite{ISO21448},  is currently available as ISO/PAS 21448 specifications with a formal release planned in 2021. 
The predecessor of these standards is the broader IEC 61508 standard~\cite{IEC61508}, which  outlines the functional safety guidelines for developing electrical/electronic/programmable electronic systems that are used to carry out safety functions~\cite{IEC61508}.

 We primarily focus on the concept phase (part 3) of the ISO 26262 standard, which outlines the derivation of FSRs and their allocation to functional architecture components.  The concept phase is executed on an item where an
 item is defined as \emph{``system 
 or combination of systems, 
 to which ISO 26262 is applied, that implements a function or part of a function at the vehicle level”}~\cite{ISO26262-2}. 

\skr{We agree that the claim "to generate FSRs from hazardous events ….;" is not appropriate especially when taken out of context. We meant the process of going from Hazardous event -> Safety goal -> ASIL allocation -> FSR. We revised the Background section to make these explicit.}

The derivation of functional safety requirements (FSRs) begins with creating hazardous events. Each hazardous event is a combination of a hazard, an operational mode, and an operational situation. An example of a hazardous event is a brake failure (hazard) in eco-driving mode (operational mode) while driving on a highway (operational situation). The operational modes and operational situations are derived from natural language descriptions of intended environments or situations where the system operates. This natural language description is referred to as scenario description or scenarios from hereon.

\skr{We revised the Background section to include an overview of the ISO 26262 concept phase including ASIL allocation to safety goals.}

To ensure safety from hazardous events, safety goals are defined. These goals are broad, presenting high-level safety requirements.  Each safety goal is allocated a score, termed Automotive Safety Integrity Level (ASIL), of A, B, C, or D, which specify the importance of achieving the goal (A for least important and  D for most important) during further stages of product development. The ASILs are calculated based on exposure, controllability, and severity of each safety goal according to the ISO 26262 guidelines~\cite{ISO26262-2}.
Each safety goal is decomposed into one or more FSRs~\cite{ISO26262-2}. Each FSR inherits the (maximum) ASIL from the safety goal(s) it is derived from.
 
 In the literature, there is little consensus on safety requirements being functional or non-functional requirements. FSRs are classified as functional requirements in the safety engineering domain.  
 However, FSRs are predominantly classified as quality requirements (non-functional requirements) in the software  architecture domain~\cite{bass2012software}.

\subsection{Safety tactics and patterns}
Architectural tactics encapsulate design decisions that can influence the behavior of a system with respect to a quality attribute~\cite{bass2012software}. 
Architectural tactics are abstract, do not impose a particular implementation structure, and can be seen as recommendations without a prescribed implementation.
On the other hand, architectural patterns   
are well-defined structured entities with a prescribed implementation that realize tactics. 
This paper employs safety tactics and patterns~\cite{wu2004safety,preschern2013building} which are architectural tactics and patterns   to address  safety.

\subsection{Architecture views} 
\skr{We agree that major challenges in cooperative driving arise at the system-level. The first phase of our two-phased method, i.e., deriving FSRs and allocating to architecture components, focuses only on the system level. In our method, the abstraction used for the allocation of FSRs, functional architecture, is at the system level.  We revised the Background section and Introduction section to make this explicit.}

 \skr{We agree that functional safety requirements are used as input/starting step to derive technical safety requirements. However, we disagree that mapping FSR to functional architecture components is illogical. According to ISO 26262:2018, “The functional safety requirements shall be allocated to the elements of the system architectural design”.  Thus each FSR can be allocated to a functional architecture component at the system level. Note that at the functional architecture level the individual architectural components are seen as black boxes and do not show how the functions are implemented. We have revised the background section to clarify these changes. }

 \begin{figure}[!ht] 
    \centering
    \includegraphics[scale=0.75]{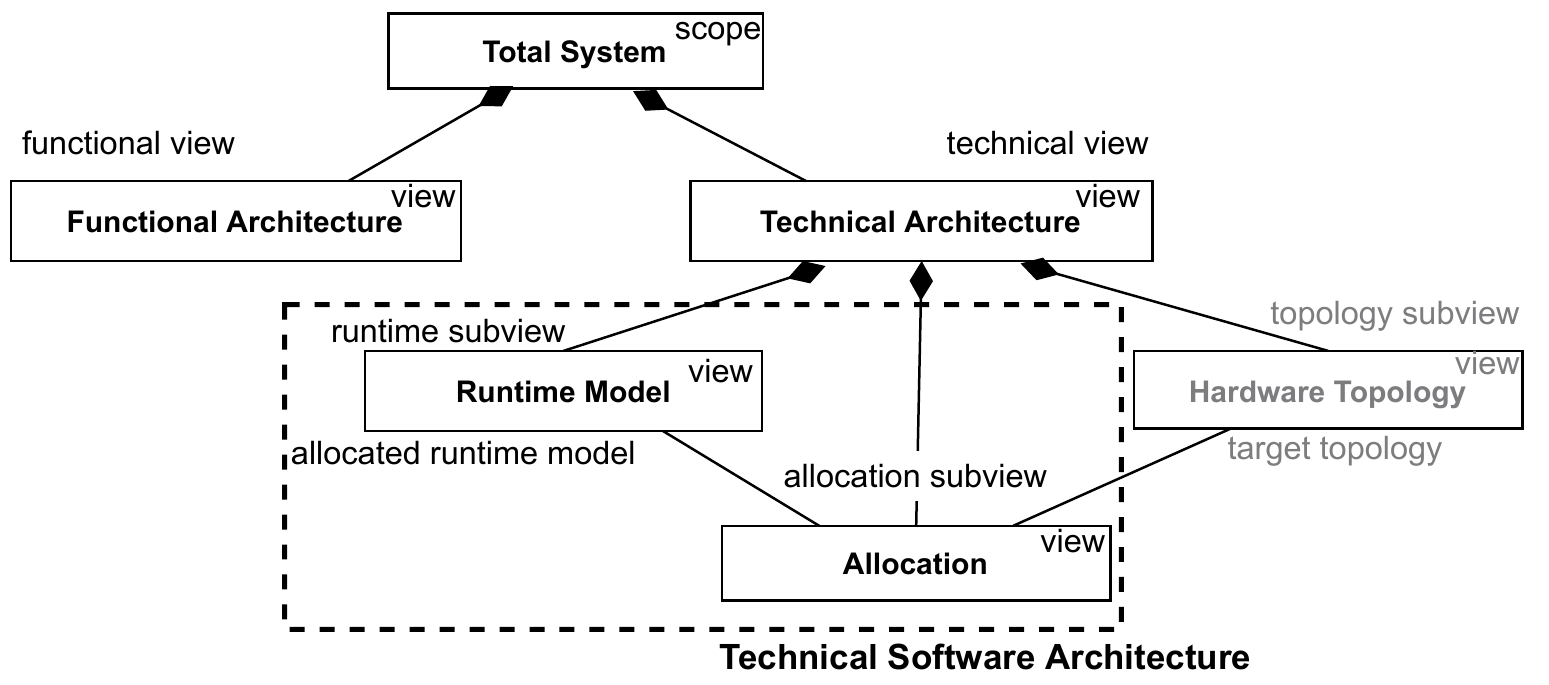}
    \caption{Functional and technical architecture views and their scope, adapted from Broy et al~\cite{broy2009automotive}. Functional and technical architecture are views of the same system at different architectural abstraction levels, with functional being the highest abstraction level. Runtime model describes system behavior while hardware topology describes the structure of hardware platform containing electronic control units, sensors, mechanical components, and the buses that interconnect them. Allocation associates elements of the runtime model with the elements of hardware topology. Runtime model and allocation together form technical software architecture.}
    \label{fig:architecture_topology}
\end{figure}
 
 We use the architecture of a system in two contexts: 
    $(i)$ to generate \fsrs from hazardous events by mapping hazardous events to
functional components of the system;
    $(ii)$ to identify whether one or more safety
tactics are used for the implementation of a functional component. 

The first context needs a functional decomposition view of the system~\cite{ISO26262-2}, known as functional architecture view~\cite{broy2009automotive,bucaioni2020technical,dajsuren2015design,staron2017automotive}. In the automotive domain, the functional architecture view outlines functional composition, functional entities, their interfaces,
interactions, inter-dependencies, behavior, and constraints in a vehicle~\cite{broy2009automotive}. This view is derived from the functional viewpoint, which considers the system from the angle of vehicular functions and their logical interactions from a black-box
perspective~\cite{broy2009automotive}. Note that the scope of this view is at the system level.

The second context demands more details that are not available in the functional architecture view but  are available in the technical architecture view (also described as the implementation view)~\cite{broy2009automotive,dajsuren2015design,Dajsuren2019,staron2017automotive}.  
The technical architecture view outlines specific software implementation, physical components (like  electronic and electrical hardware), their relationships, 
the allocation of software parts to hardware components, 
the dependencies among software and hardware components, and constraints~\cite{broy2009automotive}.  
Clearly, there is  strong 
conformity between the technical architecture view and the functional architecture view~\cite{broy2009automotive}.
A pictorial depiction of these two architectural views is shown in Figure~\ref{fig:architecture_topology}.  

We chose the technical architecture view since it enables identifying whether one or more safety tactics are implemented, and 
this view is readily available, as it is mandatory in automotive projects~\cite{broy2009automotive}. In contrast, other views 
 might lack necessary detail or may be outdated. In the rest of this paper, we discuss the runtime model and allocation part of the technical architecture view, together termed as technical software architecture.

\section{Methodology}
\label{sec:methodology}

\label{sec:sub:methodology_overview}

\skr{We address how the proposed method derives FSRs for cooperative driving and checks for the fulfillment of FSRs in the technical architecture in the respective methodology sections. We revised both the methodology parts, especially the introduction paragraph of these sections, to give a clear overview. 
}

We propose a \approach that checks whether the technical software architecture of a vehicle fulfills the FSRs for cooperative driving scenarios.
The \approach consists of two parts: $(i)$ derive FSRs for cooperative driving scenarios (see Section~\ref{sec:sub:methodology_1} and Figure~\ref{fig:RQ1}), and
$(ii)$ check whether the derived FSRs  are fulfilled in the technical software architecture of a vehicle (see Section~\ref{sec:sub:methodology_2} and Figure~\ref{fig:RQ2_RQ3}).

FSRs for cooperative driving shall be implemented in individual vehicles. 
The ISO 26262 standard recommends mapping of FSRs (or breaking down FSRs) to individual system architecture components~\cite{ISO26262-2}.
Further, such a mapping is crucial given the complexity and scale of the system.
Referring to the existing solutions from the safety engineering discipline~\cite{hommes2012review}, the current methods do not map derived FSRs for cooperative driving scenarios to individual vehicle components~\cite{kochanthara2020semi}. 
Our solution bridges this gap by integrating a cooperative functional architecture (with its individual components belonging to the vehicular functional architecture) with the existing methods to derive \fsrs. 
This step is presented in detail in Section~\ref{sec:sub:methodology_1}.


Next, we check whether the derived \fsrs are fulfilled in the technical software architecture of a vehicle. 
Our \approach of assessing  the fulfillment of derived FSRs is a combination of techniques adapted primarily from the software architecture domain. With no existing architecture assessment techniques addressing the quality attribute of functional safety in the context of automotive systems, the proposed  \approach takes inspiration from traditional architecture assessment techniques like ATAM~\cite{bass2012software,kazman1998architecture} and employs  the safety tactic framework~\cite{wu2004safety,preschern2013building,preschern2013catalog} to leverage existing architecture knowledge.
This part of our \approach is presented in Section~\ref{sec:sub:methodology_2}.

\rv{While the paper introduces the notion of architectural views and specifically focuses on functional and technical views, the differing concerns of these views never become clear in the approach.}

\skr{ Moving towards partially or full automated driving requires having an integrated view on functional safety and cybersecurity. In general, this will have tremendous impact on system and software architectures, both in individual and cooperative scenarios (see e.g. A Riel, C Kreiner, R Messnarz, A Much, An architectural approach to the integration of safety and security requirements in smart products and systems design CIRP Annals 67 (1), 173-176.
I believe that the authors should at least position their proposed method against this challenge. }

Alongside functional safety, cyber-security is another area that is increasingly addressed together with functional safety~\cite{riel2018architectural}. 
The scope of our approach is limited to functional safety and security is out of our scope.
Moreover, FSRs are often fulfilled dedicatedly in hardware or a combination of hardware and software. 
Even though the first part of our method associates FSRs to architecture components at the system level, the second part of our approach focuses on software.  
FSRs fulfilled in hardware architecture (hardware topology in Figure~\ref{fig:architecture_topology}) is beyond the scope of our method.

\subsection{Derive \fsrs for cooperative driving}
\label{sec:sub:methodology_1}



The deriving \fsrs part of our method needs only a black box view of individual vehicle functions and interactions among these functions.  Therefore, we use the functional architecture view for deriving \fsrs for cooperative driving. Note that functional architecture is the overall system architecture (see Figure~\ref{fig:architecture_topology}), which includes both hardware and software components.

\skr{We added item definition (in the context of vehicular as well as cooperative perspectives) to the methodology section.}

\begin{figure}[!ht]
    \centering
    \includegraphics[scale=0.69]{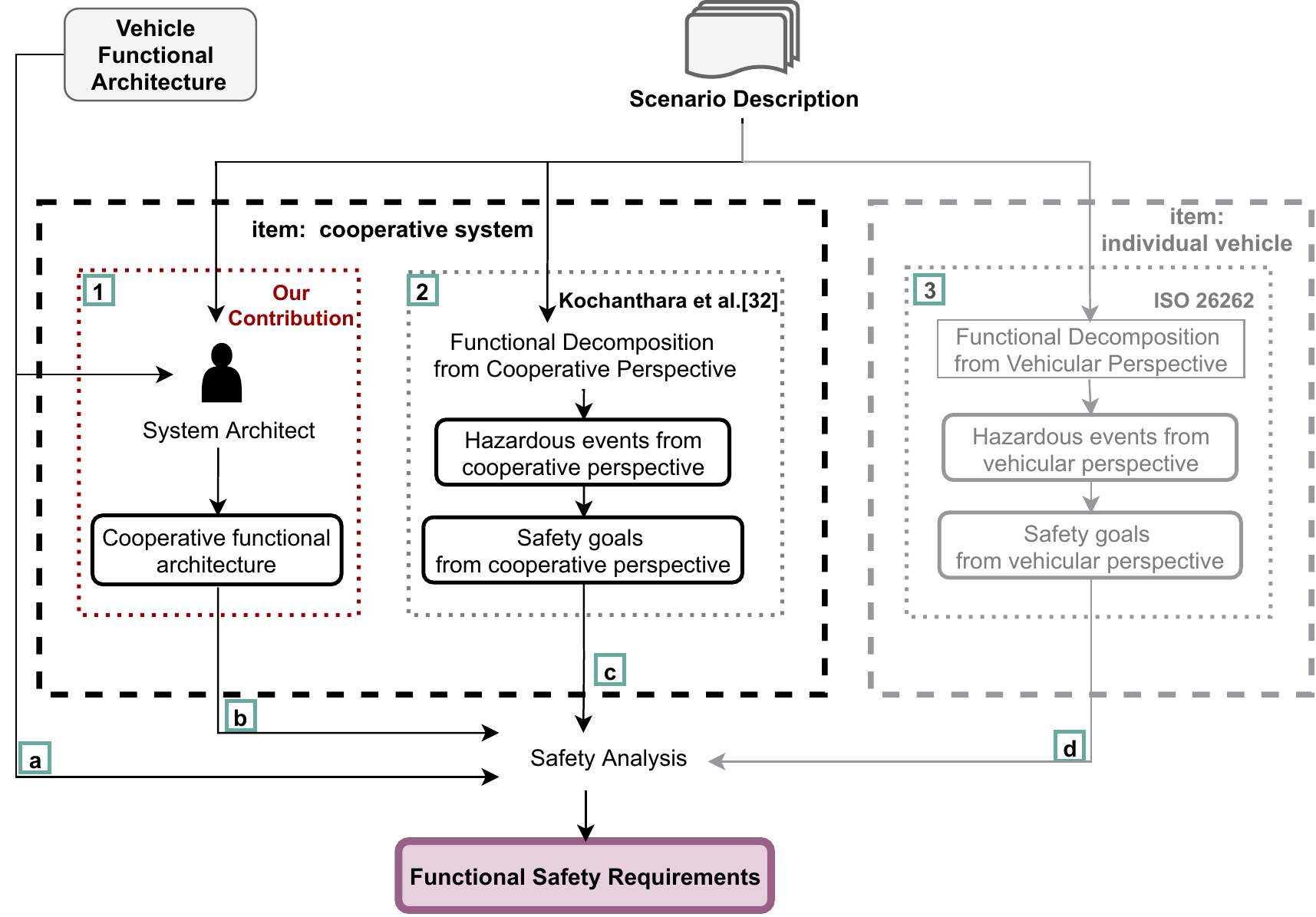}
    \caption{\Approach to derive FSRs for cooperative driving scenarios. The gray part on the right is the traditional method from ISO 26262~\cite{ISO26262-2}, and the black part on the left is our addition to the traditional approach. System architect represents external entities involved in creating the cooperative architecture.
    }
    \label{fig:RQ1}
\end{figure}

We extend the traditional \approach outlined by the ISO 26262 standard~\cite{ISO26262-2} to derive FSRs for cooperative driving scenarios. The traditional approach (the concept phase of ISO 26262) is executed on an individual vehicle as the item. We propose a similar approach to be executed on the entire cooperative system in parallel. Figure~\ref{fig:RQ1} presents an overview of the proposed \approach.  We first outline the traditional \approach, followed by our prior work on its extension~\cite{kochanthara2020semi} and our new contribution. 
For the rest of the paper, we use the term \textit{vehicular perspective} for an individual vehicle as a unit under consideration 
and \textit{cooperative perspective} for a set of vehicles as a unit under consideration.

Traditionally, \fsrs for a vehicular perspective 
 are derived  by mapping the safety goals for a vehicle on to the individual components of the vehicle's functional architecture. 
This process of mapping, also termed  safety analysis, 
captures information on the malfunctioning of a component that can lead to violation of a safety goal. Safety analysis is performed using  a systematic process like fault tree analysis (FTA)~\cite{lee1985fault} or failure mode effect analysis (FMEA)~\cite{fmea}. To conduct safety analysis, we need two inputs: $(i)$ the functional architecture that captures a system's decomposition into functional components and the interconnection between these components, and $(ii)$ safety goals.

According to ISO 26262 guidelines~\cite{ISO26262-2}, safety goals are derived from hazardous events. 
 Hazardous events are found by decomposing the scenario description using the hazard analysis and risk assessment technique (HARA)~\cite{ISO26262-2}. 
  This \approach to derive FSRs is depicted by part 3 and flows $a$ and $d$ of Figure~\ref{fig:RQ1}, with $a$ and $d$ acting as inputs to safety analysis.
 This method of deriving FSRs from scenario descriptions
 has been  standard practice in the automotive domain~\cite{ISO26262-2} for at least  a decade~\cite{ISO26262-1}.

 \skr{
We also added the step of  ASIL allocation explicitly to our method.  }
 
 During safety goal derivation using HARA, each safety goal is assigned an ASIL level. The ASIL level is allocated based on the severity of the damage possible by the hazardous event, and  the probability of exposure and  controllability of the vehicle during the event, according to the metric provided by ISO 26262~\cite{ISO26262-2}.
 Each FSR inherits the highest ASIL of the safety goal(s) it is derived from. 
 An FSR with ASIL `D' indicates that the most stringent safety measures must be applied to meet the FSR. In contrast, ASIL `A' indicates a lower risk and lower level of safety measures.


\skr{We agree that “shall” or “have to” is appropriate instead of “should”. We have revised the examples in the paper (in the  Methodology section and Case Study section) as well as our entire case study (and the entire replication package) to reflect this change.  }

In a cooperative system, a safety goal for one vehicle can lead to an \fsr in another vehicle. For example, consider a simple cooperative driving scenario of  one  vehicle (\emph{follower}) autonomously following another manually-driven vehicle (\emph{leader}) using vehicle-to-vehicle communication for  coordination. A safety goal in this setting is:  \textit {\lq\lq the follower  shall autonomously accelerate in accordance with the acceleration of the leader."}  
 Even though the safety goal seems to belong to the autonomously accelerating component of the \textit{follower}, it also maps to the functional architecture component(s) of the \textit{leader}. 
 This  safety goal leads to
 the following FSR on the acceleration sensing component of the \textit{leader}:\textit{ \lq\lq failure in the acceleration sensing component of leader shall not communicate incorrect acceleration information to the automatic steering component of the follower''}. 
 Failing to meet this 
 requirement (and its associated safety goal)  can   potentially lead to a crash.
 Such safety goals, however, will only be visible in the cooperative perspective.

\skr{We have two items: individual vehicle(s) and the cooperative system that consists of individual vehicles and other actors like the cloud. 

The proposed safety analysis takes two perspectives: vehicular and cooperative. The vehicular perspective is the same as safety analysis according to ISO 26262 where the vehicle is considered as an item and the item boundary is the same as individual vehicular functions. In the cooperative perspective, we consider the entire cooperative system and thus the item here is the entire cooperative system that contains multiple vehicles and maybe other actors like the cloud which interact with each other to enable cooperative capabilities. This can be seen as a “combination of systems” according to the item definition in ISO 26262. The cooperative perspective makes the definition of the interface between the individual actors (like individual vehicles and other actors like the cloud) explicit.  We revised the Methodology section with the explicit mention of item definition in our context.
}

In the proposed method, we have one item per individual vehicle type, and an item for the entire cooperative system of which the vehicles are part.
A cooperative system can have more than one type of vehicle (for example, two vehicles with different functional architectures forming a cooperative system) and other entities like a cloud, enabling cooperative driving capabilities. 
In the case of more than one type of vehicle (with different functional architectures), each kind of vehicle will form an item. 
For each item, except for a cooperative system, the traditional ISO 26262 analysis described above is applicable.
We believe that two items, as shown in Figure~\ref{fig:RQ1} will generalize to other scenarios that require more than two items 
since other cases only include replication of traditional ISO 26262 analysis (as shown on the right part in Figure~\ref{fig:RQ1} in gray color) for each item (i.e., each unique functional architecture). In any case, there will only be one cooperative functional architecture and thus only a cooperative item.  
For the rest of this section, we consider two items: an individual vehicle (representative of all vehicle functional architectures) and the cooperative system.

We propose that FSRs for a cooperative system  are derived from: 
 $(i)$ safety goals from the vehicular perspective (as in the traditional \approach), and 
 $(ii)$ safety goals from the cooperative perspective.
 Along these lines, our prior work~\cite{kochanthara2020semi}  extended the traditional process to derive safety goals for the vehicular perspective to the cooperative perspective (annotated as part 2 in Figure~\ref{fig:RQ1})
to cover FSRs from both perspectives.
This process partitions the scenario description into vehicle-specific and cooperation-specific parts. 
Next, we apply the traditional safety goal identification steps to the two parts.
FSRs from the vehicular perspective are then derived, as discussed above.

We observed that the cooperative functional architecture should be built using individual vehicle functional components. This will preserve the mapping between functional architecture of cooperative system and its implementation view (in the technical architecture of the vehicles).  A cooperative functional architecture is required for safety analysis techniques  like FTA~\cite{lee1985fault} to  derive FSRs, by mapping safety goals to components of functional architecture. 
We propose that the cooperative functional architecture  be built from $(i)$ the functional architecture of individual vehicles that constitute the cooperative  system and $(ii)$ the cooperative scenario description of the interaction between individual vehicles. 
With these requirements, system architects can create a functional architecture of the cooperative system such that
the individual components of the architecture are mapped onto the components of the functional architecture of vehicles. This process is labeled as part 1 in Figure~\ref{fig:RQ1}; the complete process of deriving FSRs from the cooperative perspective is shown by the labels 1, 2, $b$, and $c$.

 In summary, the presented \approach  maps each individual cooperative driving scenario to a set of \fsrs, where each \fsr is associated with at least an individual vehicle function, which in turn is associated with a functional component. Note that a one-to-one mapping is suggested for the efficiency of method and is not mandatory. Mapping an FSR to multiple functional components is unwise 
 for two reasons: (1) the responsibility is not clear, therefore implementation may go wrong; and (2) testing may not be feasible at that level and only integration testing can assess the achievement of that FSR.
 In the rest of the paper, we assume that each FSR can be mapped to a functional architecture component.

\subsection{Check fulfillment of \fsrs}
\label{sec:sub:methodology_2}

\skr{We address how the proposed method derives FSRs for cooperative driving and checks for the fulfillment of FSRs in the technical architecture in the respective methodology sections. We revised both the methodology parts, especially the introduction paragraph of these sections, to give a clear overview. 
}

\begin{figure}[!ht]
    \includegraphics[scale=0.58]{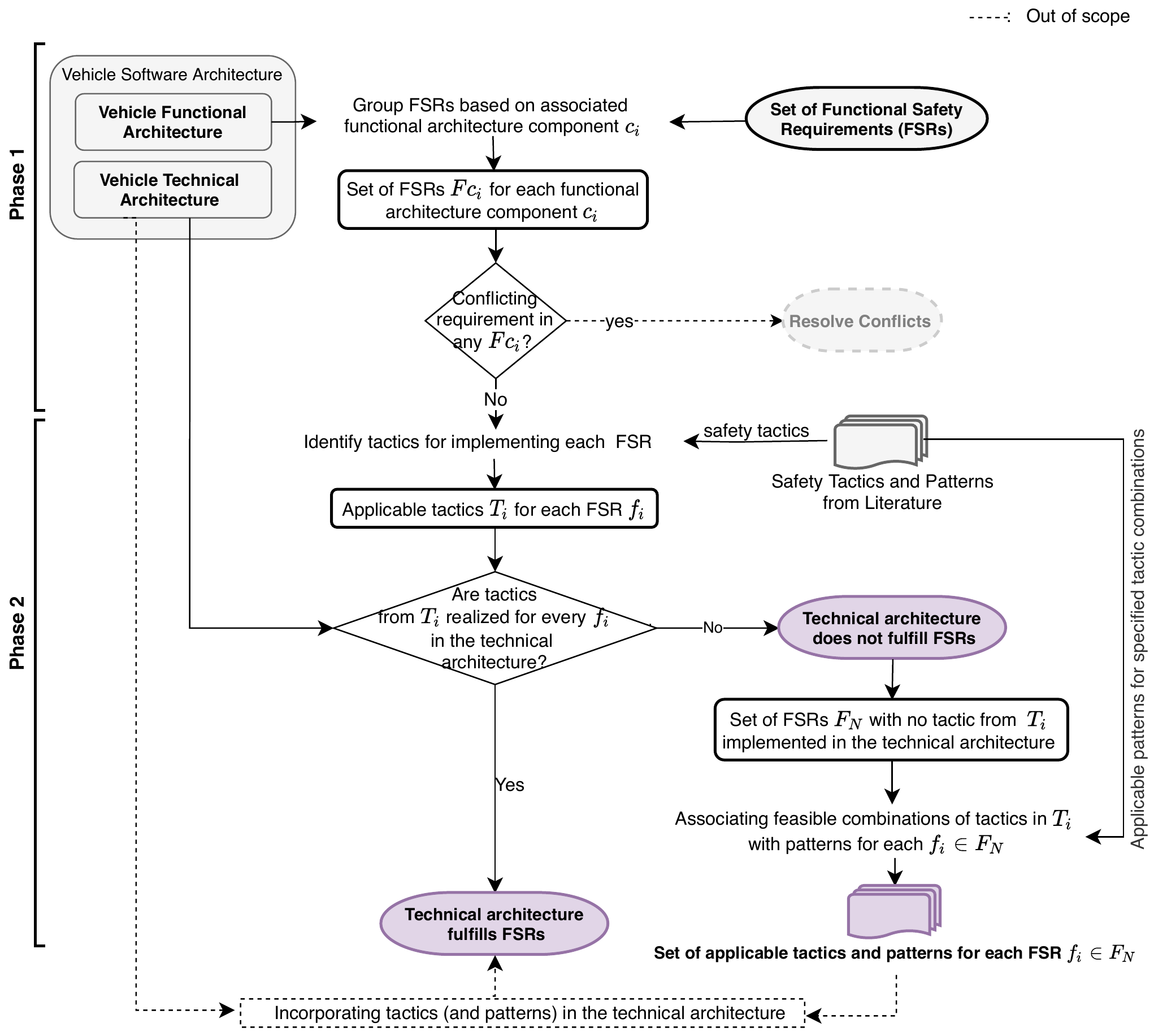}
    \caption{
    \Approach to check the fulfillment of FSRs in technical architecture
    }
    \label{fig:RQ2_RQ3}
\end{figure}

Our \approach to check for the fulfillment of FSRs in the technical software architecture of individual vehicles  is organized in two phases. Phase one ensures that it is possible to realize all the FSRs by 
identifying whether there are conflicting FSRs.
Phase two describes a systematic \approach to check for the fulfillment of FSRs in the technical architecture. Figure~\ref{fig:RQ2_RQ3} depicts an outline of the process. 

Our method uses both functional and technical views. The functional view is used for a sanity check among FSRs for conflicts. 
The technical view, in contrast, is used for checking the implementation of each vehicular function (and its associated safety mechanisms) against the corresponding FSRs. 

\skr{We agree that “shall” or “have to” is appropriate instead of “should”. We have revised the examples in the paper (in the  Methodology section and Case Study section) as well as our entire case study (and the entire replication package) to reflect this change.  }

In phase one, we check for conflicting FSRs. 
Two FSRs are conflicting if both of them cannot be fulfilled at the same time.  A hypothetical example of conflicting FSRs is: \\
\textit{FSR\_01: A failure in the actuation sensor shall be indicated by a fault message from the sensor.}\\
\textit{FSR\_02: A failure in the actuation sensor shall cease any further messages from the sensor.}\\
\textit{FSR\_01} and \textit{FSR\_02} are conflicting requirements: sending a message for \textit{FSR\_01} and not sending any message for \textit{FSR\_02} for the same event (failure in the actuation sensor), which cannot be realized simultaneously.

Comparing every pair of FSRs for conflicts will lead to a  quadratic number of comparisons (if $n$ is the number of FSRs, the number of comparisons is $n(n-1)/2 \approx O(n^2)$).
We compared \fsrs that belong to the same functional architecture component for conflicts. 
This  can reduce the number of comparisons up to a factor of $d$, where $d$ is the number of functional components (i.e., the number of comparisons can be reduced up to $n(n-d)/d \approx \Omega(n^2 / d)$). Such a reduction is possible since safety analysis techniques for deriving FSRs ensure that each FSR  belongs to only one functional component~\cite{lee1985fault,fu2018fault,ISO26262-2}.
Further, FSRs belonging to a component can have conflicts among themselves but not with the  FSRs belonging to other components.
For example, in our \cs in Section~\ref{sec:case_study}, we derived 31 FSRs across 8 functional components. 
Comparing every pair of FSRs would result in 465 comparisons; however, grouping FSRs based on functional components reduced it to 60. 
 This process is annotated as Phase 1 in Figure~\ref{fig:RQ2_RQ3}. 

The presence of conflicting requirements points to  flaw(s) in any of the following: $(i)$ the functional architecture, $(ii)$ functional decomposition of the scenario, or $(iii)$ the scenario itself. This is based on the assumption that the rest of the steps are carried out without mistakes. These conflicts need resolution before proceeding.
While resolving such conflicts is beyond the scope of this work, checking for these conflicts provides a sanity check that it is  possible to meet all \fsrs in a given technical architecture.

\rv{
\begin{itemize}
    \item So the assumption is that if there is a tactic for a FSR in the technical arch then FSR is met? What if the tactic does not fully address FSR and only to some extent? How can you make sure (e.g., using qualitative analysis techniques) that such mappings are done in a systematic way
\end{itemize}
}

An FSR may be fulfilled by a safety tactic or a combination of safety tactics.
To identify whether an FSR is fulfilled,  we propose checking the vehicle technical software architecture for the implementation of  safety tactics~\cite{preschern2013building,wu2004safety}  that can meet the FSR. 
 This is achieved in two steps: $(i)$ identify a set of safety tactics (hereafter referred to as \textit{applicable safety tactics}) such that the  implementation of  each tactic, in itself or in combination with some other tactics in the set, can fulfill the FSR; and $(ii)$ check whether any feasible combination of tactics from the applicable safety tactics that are present in the vehicle  technical architecture 
 meets the FSR.
 Note that, for an FSR $f_i$ and its corresponding functional component $c_i$, 
the  applicable safety tactics for  $f_i$ need to  be compared with only the safety tactics implementations used in the technical architecture counter part of $c_i$ and its associated safety mechanisms since $f_i$ is only associated with  $c_i$.

 
 \skr{We agree that “shall” or “have to” is appropriate instead of “should”. We have revised the examples in the paper (in the  Methodology section and Case Study section) as well as our entire case study (and the entire replication package) to reflect this change.  }

Applicable safety tactics for an FSR can be identified based on the FSR description (by navigation through a tactic hierarchy)~\cite{bass2012software,preschern2013building,wu2004safety}  or by matching the \fsr description to the descriptions of each tactic~\cite{preschern2013building}.
Consider the following example \fsr: \lq\lq \emph{failure in the acceleration sensing component of leader shall not communicate  wrong acceleration information to the automatic steering component of the follower.}''
According to the first \approach---safety tactic hierarchy~\cite{wu2004safety}---an applicable safety tactic for failure containment using redundancy is \textit{diverse redundancy}~\cite{wu2004safety}.
The same tactic can be identified by matching the \fsr description to the tactic description~\cite{preschern2013building}.
For example, the \textit{diverse redundancy} tactic's description---\textit{\lq\lq introduction of a redundant system which allows detection or masking of failures in the specification or implementation as well as random hardware failures"}~\cite{preschern2013building}---matches the FSR description.

By the end of this two step process of identifying applicable tactics and checking the technical architecture for these tactics, we will have a list of \fsrs that do not have any feasible combination of tactics implemented. 
If the list is empty, then the vehicular technical architecture fulfills all the FSRs for the given cooperative driving scenario. Otherwise, the list shows the FSRs that have not been fulfilled.

As a by-product, for each unfulfilled FSR, we will also have a set of applicable tactics such that  some feasible combinations from this set can fulfill the FSR.
These combinations point to a set of safety patterns since safety patterns are associated with the safety tactics they implement~\cite{preschern2013building}. 
These applicable safety patterns (and applicable tactics) provide the system architects with a set of possible design decisions to realize the unfulfilled FSRs.
Detailed analysis on the applicability of these safety patterns and trade-off analysis among them is beyond the scope of our work.

Note that the architecture tactics are not associated with any safety integrity level. 
 Therefore, whether a tactic can address a given ASIL level is a research topic on its own and is beyond the scope of our work. 
 Our objective for (the second phase of)  our method is to identify relevant tactics to see whether they are implemented in the technical software architecture.

\section{
Case study
}
\label{sec:case_study}

This section presents an application 
of the proposed \approach on a cooperative driving scenario: \emph{platooning}.
First, we describe the platooning scenario and the functional architecture of an individual vehicle, the two inputs to our  proposed \approach.
Next, we present the results of applying our \approach to platooning and its interpretation.
All artifacts generated are available online~\cite{kochanthara21}.

A platoon is a vehicle train in which a manually driven vehicle (referred to as \emph{leader}) is autonomously 
closely 
followed by at least one vehicle (referred to as  \emph{follower}).
In a platoon, vehicles coordinate with each other using vehicle-to-vehicle (V2V) communication.
 Platooning has shown  the potential to
$(i)$ reduce average  fuel consumption~\cite{liang2015heavy}; $(ii)$ improve safety---for example, by  preventing rear end collisions by enabling platoon-wide braking~\cite{pelliccione2020beyond}; and $(iii)$ increase traffic throughput by increasing average speed  and reducing traffic jams. In this case study, the scope of platooning is limited to highways and highway interchanges.

We applied the 
 proposed \approach on a cooperative driving software architecture developed for the i-CAVE project\footnote{\scriptsize\label{footenote:icave}\url{https://i-cave.nl/}} that is deployed on    
\emph{Renault Twizy\footnote{\scriptsize\url{https://www.renault.co.uk/electric-vehicles/twizy.html}}} -- a small electric vehicle.
The vehicle 
 is fitted with extra sensors and
actuators including a complete software stack (hereafter referred to as i-CAVE demonstrator). 
The software stack of the i-CAVE demonstrator is deployed on a combination of a real-time computer---an Advantech ARK-3520P\footnote{\scriptsize\url{https://bit.ly/AdvantechARK-3520P}}---that runs the Simulink RealTime operating system and an Nvidia's Drive PX2 platform.\footnote{\scriptsize \url{https://developer.nvidia.com/drive/}} 

\begin{figure*}[!t]
\centering
\subfloat[Vehicle functional architecture]{
\includegraphics[scale=0.49]{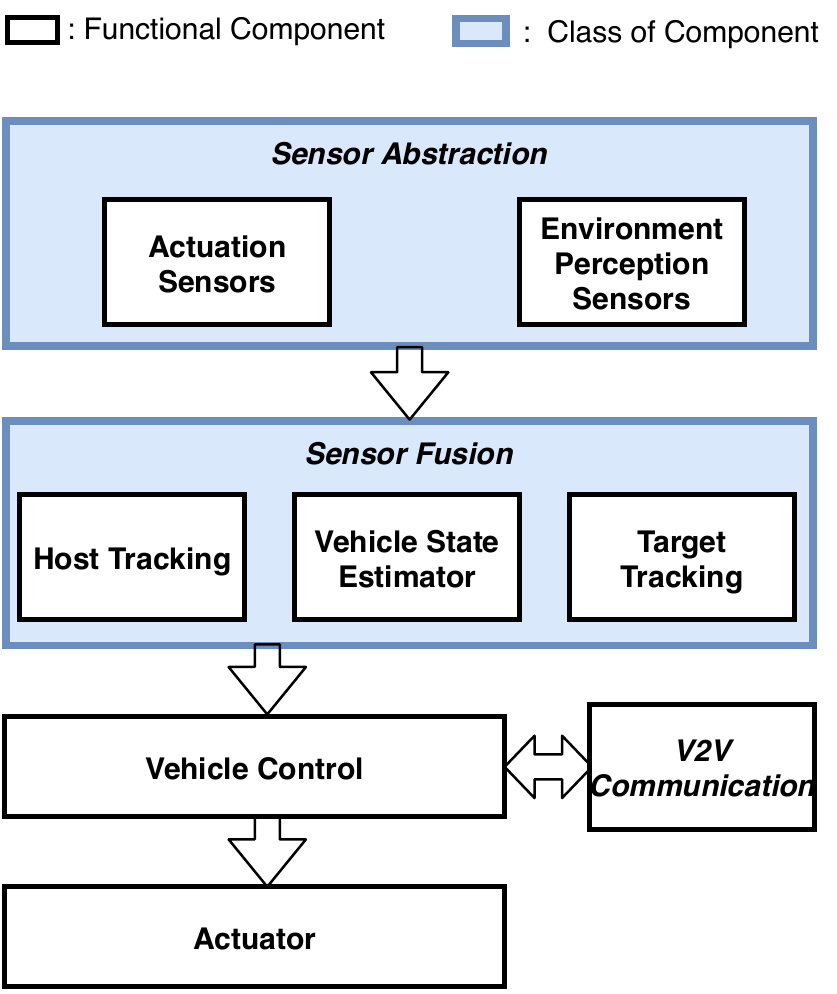}
\label{fig:vehicle_architecture}
}\qquad
\subfloat[ Cooperative functional architecture for platooning]{
\includegraphics[scale=0.47]{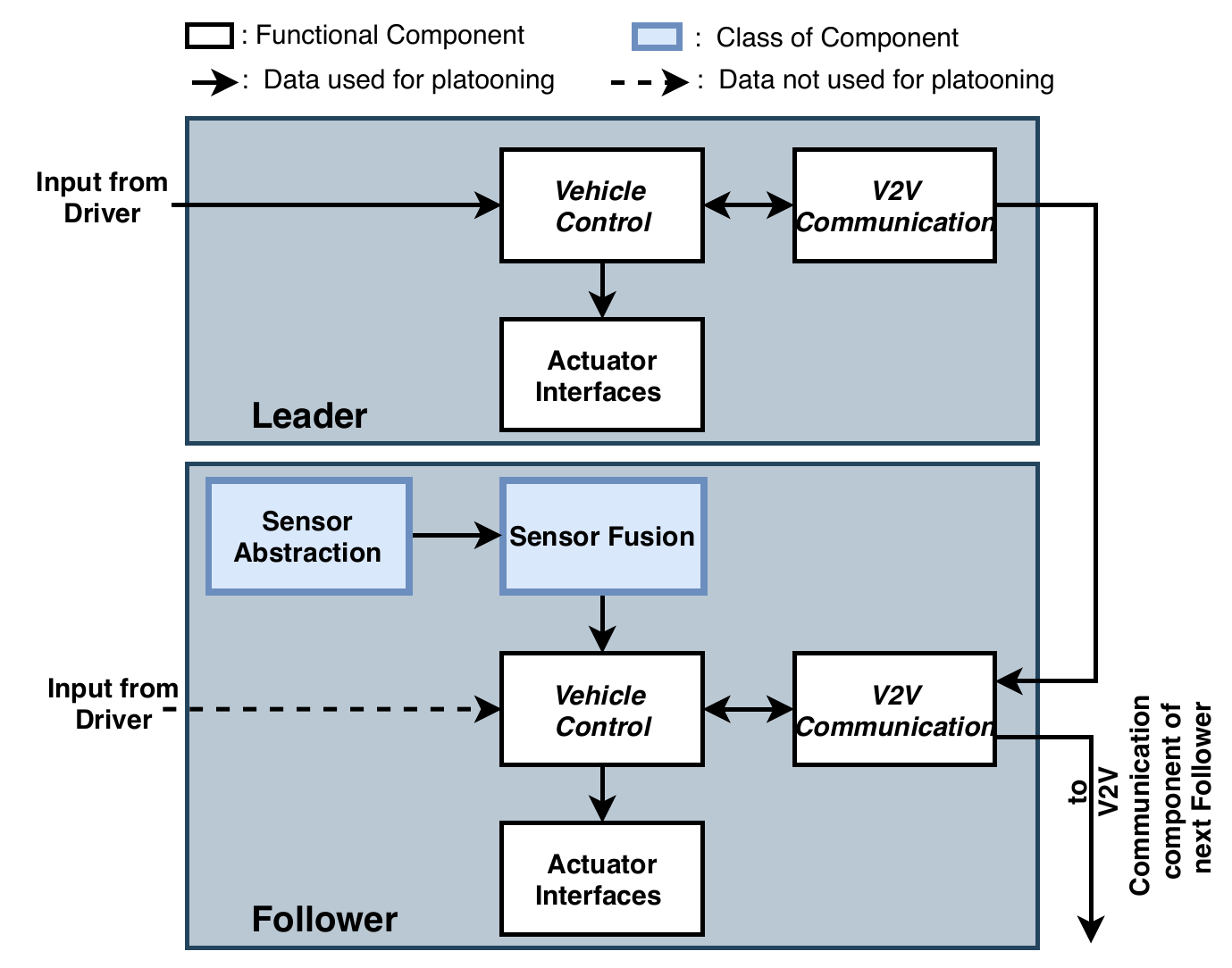}
\label{fig:platoon_architecture}
}
\caption{A simplified functional architecture of i-CAVE demonstrator and the platooning architecture (cooperative architecture) derived from it. 
    \newline
     \rv{Cooperative aspect does not become very clear. The two functional views shown in Figure 3 (one for and one cooperative?) does not reflect the defers between the cooperative and individual scenarios or solution architecture?}
    }
\label{fig:functional_arch}
\end{figure*}

A simplified functional architecture of i-CAVE demonstrator is shown in Figure~\ref{fig:vehicle_architecture}. 
For simplicity, we present only those functional components that are 
fundamental to achieve platooning. 
The arrows indicate data flow from sensor abstraction to actuator while the system as a whole is a closed control loop. 
Some of the functional architecture components are grouped to classes based on their functionality (as shown in Figure~\ref{fig:vehicle_architecture}). 
For example, sensor abstraction is a class of components that contain two types of functional components  namely actuation sensors and environment perceptions sensors.
The functional components inside each class act as independent entities and do not have data flow between them. 
The functional components of the architecture are described below: \\
    $a)$ \emph{Sensor abstraction} consists of hardware sensors and their encapsulation via its software interfaces.  
    Two classes of sensors are functionally distinguished: $(i)$ \textit{actuation sensors} that monitor vehicle state and dynamic attributes like speed and inertial measurements; $(ii)$ \textit{environment perception sensors}, like RADAR and GPS,   that monitor the vehicle's external environment and localize the vehicle on the map. 
    \\
    $b)$ \emph{Sensor fusion} combines  data from different kinds of sensors to generate information about the vehicle and its surroundings. 
    The sensor fusion of i-CAVE demonstrator has  three functional components: $(i)$  \textit{host tracking} that combines location and inertial measurement data to determine the absolute position of the vehicle, $(ii)$ \textit{vehicle state estimator} that combines acceleration information with data from actuation sensors to estimate the dynamic state of the vehicle, and $(iii)$ \textit{target tracking} component  that combines data from environment perception sensors like radar to detect objects and other vehicles in the surroundings of the vehicle. 
    \\
    $c)$ \emph{V2V communication} communicates actuation-related signals for platooning between a  vehicle and its surrounding vehicles.
    \\
    $d)$ \emph{Vehicle control} generates control signals for autonomous actuation of the vehicle using the information about the state of the vehicle, its surroundings, and information about the vehicle in front (received via V2V communication). 
    When manually driven, this component receives actuation commands from a human driver.  
    \\
    $e)$ \emph{Actuator} is hardware and corresponding software interface for accelerating, steering, and braking of the vehicle, also known as drive-by-wire interface.
\\
Note that the components 
 to fulfill non-functional requirements (outside the platooning functionality), like safety management components, 
are not shown since they are not part of
basic functional architecture needed to achieve platooning.

\subsection{Derive \fsrs for platooning}
\label{sec:sub:case_study_part_1}
Following are the steps in the first part of our \approach, depicted in Figure~\ref{fig:RQ1}. 

\emph{Functional decomposition:}
We decompose the platooning scenario description (also referred to as \emph{SD}) into five sub-scenarios. 
\begin{enumerate}[label=\emph{SC-\arabic*},leftmargin=*]
\item A vehicle can join a platoon as a follower after the last follower.
\item A follower can leave a platoon.
\item A platoon can split into two platoons.
\item Two adjacent platoons can merge into a single platoon.
\item When the leader leaves a platoon, the first follower becomes the new leader. 
\end{enumerate}
A platoon is formed when one vehicle joins another vehicle to form a two-vehicle platoon.
Eventually, a platoon is disbanded when a vehicle leaves a two-vehicle platoon. 
The join and leave actions in a platoon are performed manually by the driver of the vehicle.

The platooning scenario description is partitioned into 9  functions from the vehicle perspective and 6 functions from the cooperative perspective. 
These functions are listed in Table~\ref{tab:function_hazard}. 

\skr{P13, the term "optimal" is not commonly used in safety standards. For example, how can you define "optimal distance" from a safety perspective? An adequate replacement could be "sufficiently safe" distance.}

\emph{Hazardous events:}
Next, we identify hazards relating to these functions.
We use the seven most common guide words from the automotive domain (\textit{no, more, less, as well  as, part of, reverse,} and \textit{other than})~\cite{IEC61508} to identify 57 hazards. 
For example, the platoon function---\textit{\lq\lq keep sufficiently safe inter-vehicular distance"}---with the guide word \textit{less} creates the hazard---\textit{\lq\lq keeping less than sufficiently safe inter-vehicular distance"}---that can potentially lead to crash inside a platoon.
 The list of hazards  is  available online~\cite{kochanthara21}
 and the count of hazards derived from  each  function is shown in Table~\ref{tab:function_hazard}.

\skr{P13, the term "optimal" is not commonly used in safety standards. For example, how can you define "optimal distance" from a safety perspective? An adequate replacement could be "sufficiently safe" distance.}

These 57 hazards (26 from cooperative perspective and 31 from vehicular perspective) when combined with operational modes (7 from cooperative perspective and 6 from vehicle perspective) and operational situations (2 per perspective) resulted in 340 hazardous events,
140 from vehicle perspective and 200 from cooperative perspective.  
Note that not every combination of hazards, operational modes, and  operational situations is feasible and the infeasible combinations are not considered further.
An example of a hazardous event from cooperative perspective is: \textit{\lq\lq  keeping less than sufficiently safe inter-vehicular distance (hazard) during merge with another platoon (operational mode) on highway (operational situation)"}.

\emph{Safety goals:}
For each  hazardous event, we created a safety goal to prevent it. 
We merged similar goals in each perspective 
to have 14  and 11 safety goals 
 from vehicle and cooperative perspective respectively.
For example, the safety goal \textit{\lq\lq sufficiently safe inter-vehicular distance shall be kept regardless of the operational mode or operational situation of the platoon"} is formed by combining the goals derived from 56 hazardous events.

 \skr{For the HARA: which assumptions can be/have been made in the platooning case? What about exposure/controllability, in particular for the following vehicles?}

For ASIL allocation to safety goals, we assumed that all vehicles inside a platoon, except for the leader, cannot rely on a human driver for fallback in case of any failure. For vehicles  joining or leaving a platoon,  during the process of joining and leaving, we assume a human driver for fallback in case of failures.  We have given the lowest score for \emph{controllability} in the scenarios pertaining to follower vehicles. Since the leader is human-driven, the \emph{controllability} of the leader vehicle is assumed to be the highest. 
The highest levels are assigned to the \emph{severity} if a vehicle or platoon failure causes a crash since we assumed the speed range for highways. 
We assumed different \emph{exposure} levels based on scenarios (joining platoon, leaving platoon, splitting of a platoon, merging of two platoons, and change of leader in a platoon) and operational situation (highway or highway- interchange) with the highest exposure levels in operational scenario highway. Therefore most of the safety goals are assigned ASIL D. The detailed list of \emph{exposure, controllability,} and {severity} levels assigned and resulting ASIL for each safety goal is available online~\cite{kochanthara21}.

\begin{table}
  \scriptsize
  \centering
  \caption{Hazards for functions identified from platooning description}
    \label{tab:function_hazard}
\subfloat{%
\begin{tabular}{|x{2.9cm}|x{0.82cm}|}
\hline
  \textbf{Cooperative \newline functions}
  & \textbf{Hazard (count)}  \\ \hline
     Keep optimized \newline inter-vehicular distance within a platoon & 4 \\ \hline
                        
                        Make place for a \newline vehicle to join  &  6\\ \hline
                        
                        Merge with another \newline platoon  &  4\\ \hline
                        
                        Split into two platoons  &  4 \\ \hline
                        
                        Change leader & 4 \\ \hline
                        
                        Keep proper distance to the surrounding traffic  & 4 \\ \hline 
                        
\end{tabular}
  }
  \subfloat{%
\begin{tabular}{|x{5.4cm}|x{0.8cm}|}
\hline
  \textbf{Vehicle functions }
  & \textbf{Hazard (count)}  \\ \hline
  
                        Autonomously follow the vehicle in front (follower)
                        & 3 \\ \hline
                        
                         Keep a proper distance to the surrounding traffic as part of a platoon (follower) 
                          & 5 \\ \hline
                        
                        Leading the platoon (leader)  & 4 \\ \hline
                        
                        Take leader role of platoon  
                        & 3 \\ \hline
                        
                        Switch from  leader to follower role & 3 \\ \hline
                        
                        Join platoon (follower) & 2 \\ \hline
                        
                        Leave platoon  & 2 \\ \hline
                        
                        Timely react to the actions of surrounding \newline vehicles in a platoon & 5 \\ \hline
                        
                        Follow traffic indications, signs and rules & 4 \\ \hline
\end{tabular}}
    \hspace{.5cm}

\end{table}

 \emph{Cooperative functional architecture:} 
Figure~\ref{fig:platoon_architecture} shows
a simplified cooperative functional architecture for platooning with functional components for platooning as well as the working of vehicles within a platoon at the functional level. 
 The cooperative functional architecture is created by four system architects, who are mechanical engineers involved in the development of i-CAVE demonstrator with at least a master's degree and a minimum of two years of experience in automotive architecture development. 
 The cooperative functional architecture contains the same functional components  as the vehicular functional architecture (see 
 Figure~\ref{fig:vehicle_architecture} and Figure~\ref{fig:platoon_architecture}), but only the components that are used to accomplish the cooperative functions and their interconnections are used. 
 For example, a design choice of the system architects was to communicate the information from the vehicle control functional unit of the leader to the follower and not to communicate the sensor information between the leader and the follower. Thus,  in the leader, the \textit{sensor abstraction} and \textit{sensor fusion} class of functional components are not used for cooperative driving functions. Also, these functional components are not used for leader's own driving functions since the leader is manually driven. Therefore, in the cooperative functional architecture, in the leader block,  these components are not shown for leader (see the leader  block at the top of Figure~\ref{fig:platoon_architecture}).

 \emph{Safety analysis:} Finally, \fsrs are derived by mapping safety goals to the functional architectures using fault tree analysis (FTA)~\cite{lee1985fault}. 
The FTA generated 16 \fsrs from the vehicle perspective and 15 \fsrs from the cooperative perspective. i.e., 31 in total.  
The count of \fsrs for each functional component is presented  in Figure~\ref{fig:barchart} along with some example \fsrs in the second column of Table~\ref{tab:fsrs_in_arch}.

\color{black}
\subsubsection{
Interpretation of results
}
\label{sec:sub:discussion_case_study_1}
$\hspace{1cm}$

In our \cs, the traditional safety analysis (vehicular perspective) according to ISO 26262, resulted in 16 safety goals leading to 16 FSRs.  While the proposed extension of safety analysis resulted in 9 more safety goals and 15 more FSRs, resulting in a total of 25 safety goals and 31 FSRs. The maximum number of FSRs from the vehicular perspective is associated with  the \emph{vehicle control} component (6 FSRs), while in the context of FSRs from the cooperative perspective, it is the \emph{V2V communication} component (5 FSRs). Another interesting note is that most of the FSRs (17 out of 31; 12 from  the vehicular perspective and 5 from   the cooperative perspective) is assigned with ASIL D while only a relatively low number of safety goals (7 out of 25; 6 from  the vehicular perspective and 1 from  the cooperative perspective) as assigned with ASIL D. 
This
difference in ASILs between safety goals and FSRs is caused by the fact that most functional safety requirements are related to multiple functional safety goals, and FSRs inherit the highest
ASIL of their related safety goals.

Our count of  \fsrs (31 \fsrs from 25 safety goals in total) is low compared to industry scenarios in which a similar count of safety goals are linked to more than 100 \fsrs.
We believe that the reduced  number of \fsrs is related to the simplicity of our vehicle functional architecture.
To give perspective,  a reference architecture presented in~\cite{serban2018standard} has  39 functional components
while our simplified architecture has 8.

It is possible to have overlap of \fsrs derived from both the perspectives. 
That is, the same \fsr can be derived as a result of cooperative and vehicular perspectives.  
Our \cs, however, did not result in any overlapping \fsrs. 

\subsection{
Check fulfillment of FSRs
}

Following are the steps in second part of our \approach, depicted in Figure~\ref{fig:RQ2_RQ3}. 

\begin{figure}[!h]
    \centering
    \includegraphics[scale=0.35]{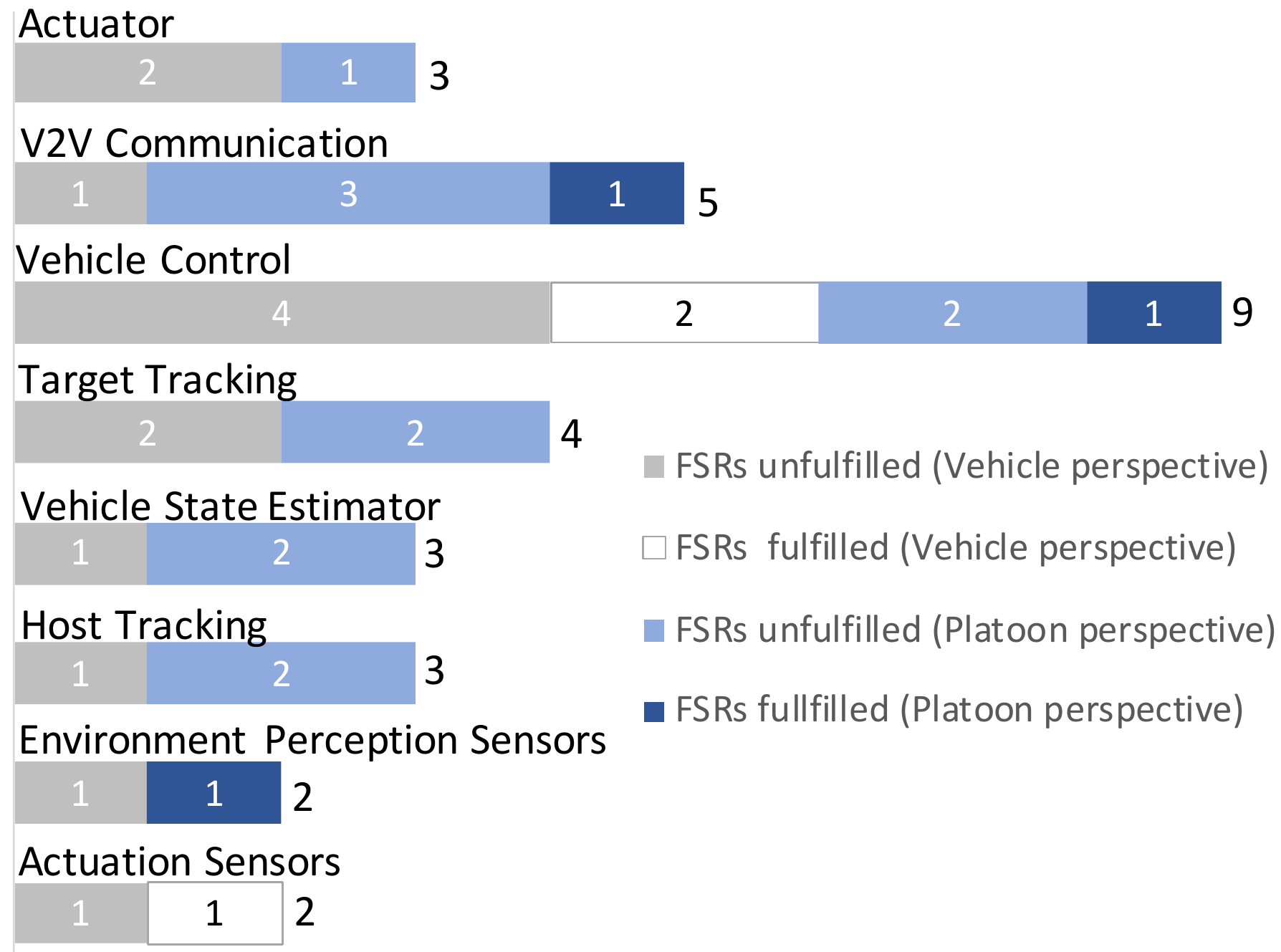}
    \caption{The 31 \fsrs, grouped by associated functional component. The total FSRs for each group is shown at the end of each stacked bar}
    \label{fig:barchart}
\end{figure}

\emph{Check for conflicts in the derived \fsrs:}
We grouped \fsrs based on their associated functional architecture component. 
For example 9 \fsrs belong to the functional architecture component \textit{vehicle control} and 3 of them is shown in Table~\ref{tab:fsrs_in_arch} (see details in the third--fifth row, first and second column).
The overall count of \fsrs grouped on 
associated component is shown in Figure~\ref{fig:barchart}. 
Within each group, we compared the descriptions of each pair of \fsrs to identify potential conflicts.
We did not find any conflict in the 8 groups.
The complete list of \fsrs grouped  by functional architecture component and compared pairwise is available online~\cite{kochanthara21}.

\emph{Identify safety tactics for implementing each FSR:}
For each \fsr we identified a list of applicable safety tactics.
We chose the following 13 safety tactics 
on which the 15 most widely used safety patterns build~\cite{preschern2013building,wu2004safety}: 
 \textit{simplicity, substitution, sanity check, condition monitoring, comparison, diverse redundancy, replication redundancy, repair, degradation, voting, override, barrier} and \textit{heartbeat}~\cite{preschern2013building,wu2004safety}.

Each safety tactic has an aim and a description of its scope~\cite{preschern2013building}. 
For example, the aim of safety tactic \textit{simplicity} is to \textit{\lq\lq avoid failure by keeping a system as simple as possible"} and its description is \textit{\lq\lq Simplicity reduces system complexity.
It includes structuring methods or cutting unnecessary functionality and organizing system elements or reducing them to their core safety functionality to eliminate hazards."}
\cite{preschern2013building}.

 \begin{table}[!ht]
\scriptsize
    \centering
    \caption{\fsrs that are found to be fulfilled in the technical architecture of i-CAVE demonstrator. \fsrs in blue cells are derived from cooperative (platooning) perspective and other \fsrs are derived from vehicle perspective. 
    }
    \label{tab:fsrs_in_arch}
    
    \begin{tabular}{|m{.9cm}|m{2.8cm}H|m{1.4cm}|HHm{5cm}| }
    \cline{1-7}
    \bfseries
& \bfseries  \fsr   & \bfseries  Applicable Tactics   &   \bfseries Applied \newline Tactics   &    \bfseries Applicable Safety Patterns   & \bfseries  Applied Safety Patterns according to Applied Tactics   &   \bfseries Implementation in  Technical Architecture  \\ \hhline{~|*6-}
\cline{1-2}

\rotatebox{90}{\makecell{Environment \\ perception \\ Sensors}}  &  \pl Failure of  \textit{environment perception sensors} shall not  result in the generation of incorrect information on distance to the surrounding vehicles and objects. 
&   "Sanity Check
Testing
Condition Monitoring
Comparison
Diverse Redundancy
Redundancy
Repair
Degradation
Voting
Masking
Override
Barrier"   &   \rotatebox{90}{\makecell{\itshape Sanity Check,\\ 
\itshape Barrier,\\ 
\itshape Heartbeat, \\
\itshape Condition \\ \itshape Monitoring}}
&   "Heterogenous Duplex Pattern
M-out-of-N Pattern
M-out-of-N-D Pattern
N-Version Programming Pattern
Acceptance Voting Pattern
N-Self Checking Programming Pattern
Sanity Check Pattern
Monitor-Actuator Pattern
Protected Single Channel Pattern"   &   /NA   &   
 Cyclic Redundancy Check (CRC) for messages \textit{(sanity check)} and  validity time per message \textit{(heartbeat)} is implemented in the  \textit{Environment perception Sensors} component while a watch dog is implemented in the safety management \textit{(condition monitoring)}. Software interface for each sensor is implemented independent of each other to protect from unintended influence between interfaces \textit{(barrier)}.
\\ \cline{1-7}

\rotatebox{90}{\makecell{Actuation \\ Sensors}} &  \ve 
External interference shall not invalidate/corrupt data from \textit{actuation sensors} .
&   "Sanity Check
Testing
Comparison
Diverse Redundancy
Redundancy
Voting
Masking
Override
Barrier
Heartbeat"   &  \rotatebox{90}{\makecell{\itshape Sanity \\ \itshape Check}}
&   "M-out-of-N Pattern
M-out-of-N-D Pattern
N-Version Programming Pattern
Acceptance Voting Pattern
N-Self Checking Programming Pattern
Sanity Check Pattern
Watchdog Pattern"   &   /NA   &  CRC and a message counter is implemented 
\\ \cline{1-7}

\rotatebox{90}{ \multirow{3}{1.5cm}{Vehicle Control}}   
&  \ve A failure in \textit{vehicle control} shall not cause generation of incorrect actuation signals  
&   "Condition Monitoring
Testing
Override
Masking
Barrier"   & \rotatebox{90}{\makecell{ \itshape 
Barrier,  \\ 
\itshape Condition \\ \itshape Monitoring}}&   Monitor-Actuator Pattern   &   /NA   &   Two independent driving modes 
are implemented in \textit{vehicle control} component.  
One mode generate  control signals (when in follower role) relying on \textit{V2V communication} and the 
other without relying on  \textit{V2V communication} \textit{(barrier)}. A monitor for checking correct working of  (and switching between) 
the two modes is implemented  
in safety management \textit{(condition monitoring)}.\\ \hhline{~|*6-}

& \pl  
A failure in \textit{vehicle control} shall neither inhibit nor modify the input from driver to further pass on.
&   "Simplicity
Sanity Check
Testing
Condition Monitoring
Comparison
Replication Redundancy
Redundancy
Barrier"   &   \rotatebox{90}{\makecell{ \itshape Simplicity}} 
&   /NA   &   /NA   &   The driver input is bypassed directly to \textit{Actuators}.   \\ \cline{2-7} 

& \ve  A failure in \textit{vehicle control} shall not cause a switch to manual drive mode while in platooning mode   
&   "Simplicity
Sanity Check
Testing
Comparison
Repair
Override
Masking
Barrier
Rollback"   &  \rotatebox{90}{\makecell{\itshape Sanity \\ \itshape Check, \\
\itshape Override, \\
\itshape Condition \\ \itshape Monitoring}}
&   Sanity Check Pattern   &   Sanity Check Pattern   &   A state machine based method for mode selection and monitoring is implemented as a part of safety management.  \\ \cline{1-2}
\hhline{~|*6-}

\rotatebox{90}{\makecell{V2V \\ Communication}}  & \pl  
Failure in \textit{V2V communication} shall not transmit incorrect information to or receive incorrect information from a vehicle joining or leaving a platoon.
&   "Simplicity
Sanity Check
Testing
Comparison
Replication Redundancy
Redundancy
Voting
Masking
Override
Heartbeat
Rollback"   &   \rotatebox{90}{\makecell{\itshape Heartbeat}}   &   "Triple Modular Redundancy Pattern
M-out-of-N Pattern
M-out-of-N-D Pattern
Sanity Check Pattern
Watchdog Pattern"   &   /NA   &  Heartbeat messages to continuously monitor reliability of communication channel are implemented in \textit{V2V Communication.}  \\
\cline{1-7}

\end{tabular}

\end{table}

 \emph{Applicable tactics for each FSR:}
 To identify whether an implementation of a safety tactic can realize an \fsr, the aim and description of the safety tactic is matched with the description of the \fsr. 
 Examples of \fsrs, safety tactics that match them as well as their implementation are presented in Table~\ref{tab:fsrs_in_arch}.
Table~\ref{tab:fsrs_in_arch} also shows that the first \fsr listed does not match the simplicity tactic (not present in column 3) resulting from the inherent complexity of \textit{environment perception sensors}.
 A complete list of the 31 \fsrs and matched safety tactics is available online~\cite{kochanthara21}.
 
\emph{Check for safety tactics implementations in technical architecture:}
Finally, to identify whether the vehicle architecture meets an \fsr, we analyzed the implementation of the associated functional architecture component in the technical architecture of the i-CAVE demonstrator.
The technical architecture of the i-CAVE demonstrator is implemented  in MATLAB/Simulink. 
We inspected the MATLAB code as well as the Simulink state flow diagram to identify functional architecture components 
as well as any associated safety management system.
We mapped the implementations of these functional architecture components to the safety tactics identified for each FSR
to evaluate whether each \fsr is fulfilled by the technical architecture. Table~\ref{tab:fsrs_in_arch} shows  
the FSRs that are found to be fulfilled in the technical architecture, the tactics applied  from the set of applicable tactics, and how the specific combination of applied tactics fulfills the corresponding FSR. Also, an example of FSR that is found to be unfulfilled is:
 \emph{``A failure in \emph{Actuator} (software interface) should not cause 
propagation of incorrect control signals to hardware actuators".} 
A complete list of unfulfilled FSRs is available online~\cite{kochanthara21}.
 
For each functional component, Figure~\ref{fig:barchart} shows the count of \fsrs that are realized and not realized from the  vehicle as well as the platooning perspective, respectively. 
Recall from  Section~\ref{sec:sub:case_study_part_1} that we derived 16 and 15 FSRs from the vehicle and platooning perspective, a majority of them relate to vehicle control. 
Out of the 16 \fsrs for the vehicle perspective, 3 \fsr are fulfilled by the vehicular technical architecture and the remaining 13 \fsrs are unfulfilled.
Likewise, for the cooperative perspective, 3 \fsrs are fulfilled and the remaining 12 \fsrs are unfulfilled. 
We showed our results to the four system architects of the i-CAVE project. They confirmed that fulfilled \fsrs are implemented and unfulfilled \fsrs are not implemented in i-CAVE demonstrator.  
For the 25 \fsrs unfulfilled by i-CAVE demonstrator, we provide a list of applicable safety patterns that can act as a starting point for the next design iteration of the technical architecture.

\subsubsection{Interpretation of results
}
$\hspace{1cm}$

\label{sec:sub:discussion_case_study_2}

Our study shows that the technical architecture meets only 6 out of 31 FSRs (with all six fulfilled \fsrs presented in Table~\ref{tab:fsrs_in_arch}). 
In our \cs, we checked whether FSRs are fulfilled in the  i-CAVE demonstrator using 13 safety tactics.  The set of tactics was  chosen based on their use in the 15 most widely used safety patterns~\cite{preschern2013building}. It is possible that we might have classified some fulfilled FSRs to be unfulfilled since we considered only 13 tactics. However,   the architects of the i-CAVE project agreed to our findings. This indicates that our classification was correct.

Our assessment using these safety tactics showed that, out of the 15 \fsrs from the cooperative perspective, 12  were unfulfilled.   
Notably, we found almost as many unfulfilled \fsrs from the vehicle perspective as from the cooperative perspective.
An explanation for this observation relates to the capabilities of the vehicle behind the i-CAVE demonstrator. 
The i-CAVE demonstrator uses a Renault Twizy\footnote{ \scriptsize https://www.renault.co.uk/electric-vehicles/twizy/specifications.html} which is a bare-bones two seater electric vehicle. 
To give perspective, the Renault Twizy is small enough ($2338mm \times 1381mm$) to be used in bicycle lanes, is lightweight (gross weight of 690 kilograms) and has a driving range of up to 51 kilometers. 
In contrast, Tesla's entry level vehicle, Model 3,\footnote{ \scriptsize https://www.tesla.com/model3} is almost double in dimension, 
three times in gross weight, 
and more than ten times in driving range. 
As a result, the i-CAVE demonstrator has limited features and components.
Also, the  demonstrator is a work-in-progress being developed iteratively by a multi-domain team. 
Since some parts of the technical architecture were not implemented during our \cs, our results 
 merely point to the missing implementations as unfulfilled FSRs in the vehicle perspective.
Future iterations of the demonstrator\footnote{\scriptsize\url{https://i-cave.nl/}} can use this list of unfulfilled \fsrs from both the vehicle and the cooperative perspective to improve the i-CAVE technical architecture.

\skr{P14. "Clearly, replications are required to verify generalizability and scalability of our method."  = " … the verify generalizability…"}

In summary,  our \cs found unfulfilled FSRs from the cooperative perspective showing the viability and applicability of the proposed \approach. 
The results of our case study show  better coverage of safety goals by providing additional FSRs as compared to the ISO 26262 process~\cite{ISO26262-2}.
The current safety engineering  methods to derive \fsrs are outlined by ISO 26262 standard~\cite{ISO26262-2,nilsson2013functional} which lacks the cooperative perspective.
It provided valuable insights in the context of i-CAVE project. 
In our \cs,
we found 15 \fsrs from the cooperative perspective, making up 48\% of all \fsrs.
Yet it is still a mere illustration of our \approach. 
Clearly,  replications are required to verify the generalizability and scalability of our \approach.
Nonetheless, our results corroborate 
the existing body of knowledge~\cite{dajsuren2019safety,nilsson2013functional,saberi2018functional} in showing that the current safety standard misses \fsrs from the cooperative perspective.

The results of our case study show  better coverage of safety goals by providing additional FSRs as compared to the ISO 26262 process~\cite{ISO26262-2}.
The current safety engineering  methods to derive \fsrs are outlined by ISO 26262 standard~\cite{ISO26262-2,nilsson2013functional} which lacks the cooperative perspective. In our \cs,
we found 15 \fsrs from the cooperative perspective, making up 48\% of all \fsrs.
Our results corroborate 
the existing body of knowledge~\cite{dajsuren2019safety,nilsson2013functional,saberi2018functional} in showing that the current safety standard misses \fsrs from the cooperative perspective.

\section{Discussion}
\label{sec:discussion}

We present a deeper exploration into the proposed method in terms of implicit assumptions. 
We describe how our solution is likely to apply to cooperative driving scenarios in real-life. 
Below we discuss the implicit assumptions in our method, the scalability, generalizability, and the scope of our method.

\subsection{Assumptions}
The proposed method borrows some assumptions applicable to single-vehicle and applies them to cooperative driving.
These assumptions are derived from the safety engineering domain as well as the software architecture domain.
For example, it is assumed that  proper functional separation of the system is always possible. This results in every FSR being mapped to exactly one functional component.
Such a functional separation is a standard practice in the safety engineering domain and has been followed for at least five decades~\cite{lee1985fault}.
This separation is also underlined by  the 
product development standard in the automotive domain---ISO26262~\cite{ISO26262-1,ISO26262-2} and   automotive architecture frameworks~\cite{broy2009automotive}.
Nonetheless, the applicability of this assumption in cooperative driving is not established.

Similarly, the second part of our \approach relies on two assumptions:
$(i)$ it is possible to map functional components to implementations in the technical software architecture; and 
$(ii)$ every FSR can be fulfilled by a combination of safety tactics.
Our first assumption comes from the architecture frameworks in the automotive domain~\cite{broy2009automotive}. 
The assumption about safety tactics stems from the architecture domain, which considers safety tactics as design primitives, and architectures are formed by the combination of design primitives~\cite{bass2012software}.
Mature architecture assessment methods like ATAM also rely on this assumption, albeit in the context of tactics for the quality attributes they focus on~\cite{kazman1998architecture,bass2012software}.
Nonetheless, the applicability of this assumption in the automotive domain is not established.

\subsection{
Applicability}
\skr{I understand (and agree) that the selected case is not really appropriate if the proposed methods can scale adequately to industrial projects. However, I encourage the authors to elaborate arguments on the subjects of suitability and scalability based on the key properties of their proposed approach. }

Our case study presents a simplified version of real-life cooperative driving use-cases. 
In real-life, the proposed method should work on a bigger scale and apply to  cooperative driving systems with  various entities.
The potential factors limiting the scalability and generalizability of a method include the complexity of the system, heterogeneity of participating systems (e.g., different types of vehicles (car and truck) and/or vehicles from different manufacturers), and inclusion of entities other than participating vehicles, like the cloud, to enable cooperative driving functionalities. 

\emph{Scalability:} Our method is modular, which means that it is likely to scale to complex systems.
The method uses two levels of abstraction at the architecture level. 
The functional architecture view separates functionalities such that each component performs one unique function and collectively performs a cooperative driving function. 
This ensures that the safety requirements for cooperative driving functions can be allocated to individual vehicular components without entering into their implementation details. 
In the second part of the method, all the FSRs pertaining to one component are assessed against their implementation details. 
Segregation of safety requirements pertaining to each component and handling each component separately, ensures the applicability of our approach to complex systems.

\emph{Heterogeneity:} The functional architecture view acts as a black box separating functional components and interaction among functional components from their implementation, making our approach agnostic of vehicle type and brand. 
As a result, we believe that our approach can assess the safety of cooperative driving systems that involves different kinds of vehicles (for example, platoon containing both trucks and cars) as well as different automotive brands (for example, platoon containing cars from BMW and GM) as long as the functional architectures of the participating entities are provided.

\emph{Entities other than vehicles:}
Entities enabling cooperative driving functionalities can be beyond participating vehicles. 
One such example  is cloud communication. 
The cooperative architecture and corresponding item definition in the first phase of our approach are specifically introduced to ensure that all the entities involved in enabling cooperative functionalities are systematically considered in the safety analysis. 
For example, in the use cases that include cloud communication, the cloud will be a part of the cooperative architecture.


\emph{Fail-operational and fail-safe designs:} Our work is designed for cooperative systems irrespective of their operational design domain and whether they are designed to be fail-operational or fail-safe~\cite{hasan2020fail,sawade2018robust}. 
The proposed method is generic for both fail-operational as well as fail-safe systems. 
Our method ensures that both cooperative and vehicular perspectives are covered while deriving FSRs.

\section{Related work}
\label{sec:related_work}

The proposed method has two parts: (i) deriving FSRs for cooperative driving and (ii) check whether each FSR is fulfilled in the technical software architecture of the vehicle. These two aspects are addressed separately in the literature,   and the related research primarily stems from two domains: software architecture and safety engineering.

\subsection{Software architecture}
A variety of architecture assessment techniques has emerged from the software architecture research community  in the past three decades. These architecture assessment techniques assess the ``goodness"~\cite{bass2012software} of an architecture(s) with respect to some property (or a set of properties). Such properties are  termed as quality attributes. 
Quality attributes can be divided into two broad categories: operational (e.g. reliability, performance)   and development (e.g. maintainability, re-usability)~\cite{bosch1999software}. This paper focuses on  \emph{functional safety} as an operational quality attribute.

The software architecture assessment techniques proposed for operational quality attributes~\cite{babar2004framework,dobrica2002survey}  mainly use mathematical  modeling \& analysis  and  scenarios of system operation (also known as scenario-based techniques) to uncover whether the architecture achieves the intended quality attributes sufficiently~\cite{bengtsson2004architecture}.
The assessment techniques that use mathematical modeling mainly focus on reliability and performance as quality attributes~\cite{roy2008methods}. Some studies have shown that these techniques are not scalable and hence not suitable for complex systems of systems like cooperative driving systems that are built by multiple inter-disciplinary teams~\cite{roy2008methods}. 

The prominent scenario-based architecture assessment methods for operational quality attributes are the Architecture Trade-off Analysis Method (ATAM) \cite{kazman1998architecture,bass2012software}, Scenario-Based software Architecture Re-engineering (SBAR)~\cite{bengtsson1998scenario}, Software architecture Comparison Analysis Method (SCAM)~\cite{stoermer2003scam}, Domain Specific software Architecture comparison Model (DoSAM)~\cite{bergner2005dosam}, and Pattern-Based Architecture Reviews (PBAR)~\cite{harrison2010pattern}.

PBAR is designed for light weight evaluation, primarily performed on small projects with some case studies on projects with at most 10 developers~\cite{harrison2010pattern,harrison2013using}.  
It is not suitable for evaluation of complex safety-critical systems that we assess in this paper~\cite{harrison2010pattern}. 
SCAM and DoSAM are designed for comparing different architectures rather than assessing an individual  architecture~\cite{stoermer2003scam,bergner2005dosam}. These methods grades each of the architectures under comparison on a normalized scale, typically from 0 to 100, and use this to characterize the fitness of a candidate architecture in contrast to others.
SBAR is an iterative method for re-engineering of architectures for functionality based re-design~\cite{bengtsson1998scenario} including for architectures that might not properly separate functional concerns. 
We assume that the automotive architectures for our analysis are designed based on separation of functional concerns since 
this is a standard practice in the automotive domain, enforced by safety engineering~\cite{ISO26262-1,ISO26262-2}.
Moreover, SBAR suggests scenario-based techniques for development quality attributes and simulation based assessment for operational quality attributes~\cite{dobrica2002survey}. In contrast we consider scenario-based methods for the operational quality attribute \textit{functional safety}.  

ATAM is the most mature and widely used architecture assessment method in practice~\cite{bass2012software}. 
ATAM, in its current form, is primarily used to analyze trade off among different quality attributes and to identify stress points and sensitivity points in the architecture under assessment. ATAM facilitates usage of existing knowledge in the form of tactics, which we take inspiration from and reuse in our proposed method.

ATAM considers six quality attributes, however, functional safety is not one of them~\cite{bass2012software}. Note that case studies of ATAM's application to safety critical domains like avionic systems~\cite{barbacci2003using} do not stress safety as a primary quality attribute either.
Even though ATAM provides some methods for scenario elicitation, it does not provide a systematic method for scenario decomposition to generate requirements for individual architecture components. This is crucial in the context of systems of systems since a scenario may lead to a multitude of requirements affecting different systems which are to interact with each other to perform the intended action(s).

\skr{Another thing, the authors should be careful about their claims such as "In summary, to the best of our knowledge, none of the existing software architecture assessment methods either are designed or have been used for assessing functional safety." If this claim is accurate, how do we have safety-critical systems?}

In summary, within the field of software architecture, none of the software architecture assessment methods that we found are applicable
for analyzing the functional safety of cooperative
automotive systems. 

\subsection{Safety engineering}
Now, we present related research on the application of safety engineering concepts in automotive software and system evaluation.
We primarily present related research on $(i)$ identifying \fsrs in automotive settings and $(ii)$ methods to
check 
or ensure that a technical architecture realizes \fsrs.

\skr{P6, "For instance, [6] presents" = "Beckers et al. [6] presents". There are many similar cases.}

\emph{Identifying \fsrs:}
Studies on deriving \fsrs largely focused on the perspective of individual vehicle  as a system while just a few explored the  perspective of a set of vehicles as a system.
Studies to derive \fsrs from the vehicle perspective present different mechanisms to generate safety goals and map these to the functional architecture using safety analysis methods.
For instance, Beckers et al.~\cite{beckers2014systematic} presents a model-based method to define \fsrs given safety goals while  Abdulkhaleq et al.~\cite{abdulkhaleq2017using} uses system theory for safety analysis. 

\skr{P6, "For example, [36] uses system theory for the safety analysis from the perspective of set of vehicles as a system. Another study proposed an alternative safety analysis technique, using possible accidents as a starting point to identify FSRs [36]."  This is the same study}

Studies on cooperative driving systems try to replicate the mechanisms from an individual vehicle perspective.
For example, Oscarsson et al.~\cite{oscarsson2016applying} uses system theory for the safety analysis from the perspective of set of vehicles as a system.
Another study proposed an alternative safety analysis technique, using possible accidents as a starting point to identify FSRs~\cite{stoltz2019stpa}.
Our study closely follows the study by Saberi et al.~\cite{saberi2018functional
} in deriving FSRs from cooperative driving scenarios.

\emph{
Checking or ensuring architecture realizes \fsrs:} 
Studies on a single vehicle perspective use many different approaches to ensure that systems satisfy \fsrs.
One approach uses an architecture description language for safety verification~\cite{cuenot2014applying}.
Martin et al.~\cite{martin2020combined}  uses architecture patterns to incorporate FSRs in the design phase.
Sljivo et al.~\cite{sljivo2020guiding} presents a methodology for fulfillment of FSRs at design time using design patterns and contracts.
Other approaches use formal methods to verify that systems satisfy FSRs, although the solutions do not scale~\cite{althoff2014online,bhatti2016unified,mallozzi2016formal}.

Ensuring fulfillment of FSRs as part of a cooperative system is challenging~\cite{pelliccione2020beyond}.
A majority of works on cooperative systems proposes a reference architecture from a system of systems viewpoint~\cite{pelliccione2020beyond}.  
Some other solutions look at specific architecture components in specific cooperative scenarios, thereby missing high-level insights~\cite{dajsuren2019safety}.

To the best of our knowledge these studies, with their scope of complete system architecture, focus on fulfilling FSRs during the design phase. This paper, in contrast, focuses on checking for the fulfillment of FSRs on existing architectures or when designing architectures.

\section{ Threats to validity  }
\label{sec:threats_to_validity}

Our proposed method and the findings from the case study are susceptible to threats relating to human participation and choice of techniques. 
Below, we present potential threats and our attempts at mitigating them. 

\emph{Cognitive bias:} 
Several steps of our proposed method and the related \cs rely on the expert opinions of  architects. 
This step may have resulted in cognitive bias~\cite{zalewski2017cognitive} in relation to human judgment.
To mitigate this threat, for every step that required human judgment, we consulted at least three experts (in addition to the first two authors), who performed the steps independently. 
For example, 
$(i)$ the cooperative functional architecture was created from the vehicle architecture and scenario descriptions, in consultation with four expert system architects, independently
$(ii)$ two of the authors independently checked for conflicts among FSRs; and
$(iii)$ The validity of the safety goals depends on the decomposition of a scenario description to functions. 
The decomposition of the scenario description to functions 
was validated by the third author, who is an  expert in functional decomposition with over five years of industry experience in the functional safety domain and a participant in the development of the automotive industry's functional safety standards ISO 26262 and ISO 21448. 

\emph{Technical bias:} To generate FSRs, we chose fault tree analysis~\cite{lee1985fault} as the safety analysis technique.
The choice of other techniques, like failure mode effect analysis~\cite{fmea}, may influence the outcome. 
We need empirical studies to check whether the choice of safety analysis technique introduces differences in findings.

\emph{Choice of safety tactics:} In our \cs, we checked whether the FSRs are fulfilled in the i-CAVE demonstrator using 13 safety tactics.  
The safety tactics are chosen based on their use in the 15 most widely used safety patterns~\cite{preschern2013catalog,preschern2013building}. 
This list of safety tactics is not complete and defines the scope of our case study.

\section{Future work}
\label{sec:future_work}

Our work is an initial step in the direction of functional safety assessment for   cooperative driving. 
This section presents potential future directions.

\emph{Cyber-security alongside functional safety:} Cyber-security is a prominent directions to explore in cooperative driving alongside functional safety. The connected nature of cooperative driving increases the potential attack surfaces and can compromise the system's functional safety.  Integral approaches that consider safety and security together are a potential future research direction.

\emph{Hardware topology:} Functional safety is often achieved via hardware architecture or a combination of hardware and software. The second part of our method focused only at the software level. Extending the second part of the approach to address functional safety requirements that are fulfilled specifically in hardware topology and the combination of hardware and software is the logical next step of the proposed approach.

\emph{ASILs for safety tactics:} Currently, safety tactics are not associated with ASIL levels. This means that, from the current taxonomy of safety tactics, we can only conclude whether a tactic addresses an FSR rather than whether it addresses the FSR at the specific level of ASIL. Augmenting safety tactics with ASILs is a potential future research direction. This will allow prioritizing FSRs based on the risk associated with them. This may also be a step towards a trade-off analysis where each FSR can be traded off with other requirements based on the risk associated.

\skr{P4, "It can be argued that a higher level of architecture abstraction than the technical architecture view can also be used to identify the implemented safety tactics. We leave this decision to the projects' designers. They can use the proposed method by replacing the technical architecture view as used in our method by an appropriate abstraction." Please rewrite this paragraph, it is not clear}

\emph{Alternative architecture abstraction levels:} Currently, the second part of our approach, checking fulfillment of FSRs, uses the technical architecture view. 
It can be argued that a higher level of architecture abstraction than the technical architecture view can be used instead. We plan to evaluate this in the future.

 Finally, our \approach adapts existing solutions for addressing functional safety in the context of cooperative driving scenarios. 
 Alternative \approaches to check for unfulfilled FSRs will be an interesting direction to explore.

\section{Conclusion}
\label{sec:conclusion}

This paper investigated whether the architecture of a single vehicle meets the functional safety requirements for cooperative driving. We proposed a 
 \approach to ensure that an automotive architecture is functionally safe to operate in given scenarios.
The proposed \approach derives functional safety requirements for a cooperative driving scenario and checks whether they are fulfilled in the technical architecture of a vehicle. The \approach is a combination of methods adapted from the safety engineering and software architecture domains.
We show the usability of our \approach for a cooperative driving scenario,  platooning,  on a real-life academic prototype, 
resulted in uncovering functional safety requirements that were not fulfilled by the software architecture.
Our \approach is motivated by and reinforces the notion that functional safety should not be an afterthought in the design of automotive architectures rather be used for defining the architecture of the automotive system.

%
%
\bibliographystyle{splncs04}
\bibliography{relatedwork}

\begin{thebibliography}{10}
\providecommand{\url}[1]{\texttt{#1}}
\providecommand{\urlprefix}{URL }
\providecommand{\doi}[1]{https://doi.org/#1}

\bibitem{abdulkhaleq2017using}
Abdulkhaleq, A., Wagner, S., Lammering, D., Boehmert, H., Blueher, P.: Using
  {STPA} in compliance with {ISO} 26262 for developing a safe architecture for
  fully automated vehicles. arXiv preprint arXiv:1703.03657  (2017)

\bibitem{althoff2014online}
Althoff, M., Dolan, J.M.: Online verification of automated road vehicles using
  reachability analysis. IEEE Transactions on Robotics  \textbf{30}(4) (2014)

\bibitem{alvarez2017considering}
Alvarez, P., Lerga, I., Serrano, A., Faulin, J.: Considering congestion costs
  and driver behaviour into route optimisation algorithms in smart cities. In:
  International Conference on Smart Cities. pp. 39--50. Springer (2017)

\bibitem{babar2004framework}
Babar, M.A., Zhu, L., Jeffery, R.: A framework for classifying and comparing
  software architecture evaluation methods. In: 2004 Australian Software
  Engineering Conference. Proceedings. IEEE (2004)

\bibitem{barbacci2003using}
Barbacci, M., Clements, P.C., Lattanze, A., Northrop, L., Wood, W.: Using the
  architecture tradeoff analysis method (atam) to evaluate the software
  architecture for a product line of avionics systems: A case study. Tech.
  rep., Carnegie-mellon Univ Pittsburgh PA Software Engineering Inst (2003)

\bibitem{bass2012software}
Bass, L., Clements, P., Kazman, R.: Software architecture in practice.
  Addison-Wesley Professional (2012)

\bibitem{beckers2014systematic}
Beckers, K., C{\^o}t{\'e}, I., Frese, T., Hatebur, D., Heisel, M.: Systematic
  derivation of functional safety requirements for automotive systems. In:
  (SafeComp). Springer (2014)

\bibitem{bengtsson1998scenario}
Bengtsson, P., Bosch, J.: Scenario-based software architecture reengineering.
  In: Proceedings. Fifth International Conference on Software Reuse. IEEE
  (1998)

\bibitem{bengtsson2004architecture}
Bengtsson, P., Lassing, N., Bosch, J., van Vliet, H.: Architecture-level
  modifiability analysis (alma). Journal of Systems and Software
  \textbf{69}(1-2) (2004)

\bibitem{bergner2005dosam}
Bergner, K., Rausch, A., Sihling, M., Ternit{\'e}, T.: Dosam--domain-specific
  software architecture comparison model. In: Quality of Software Architectures
  and Software Quality. Springer (2005)

\bibitem{bhatti2016unified}
Bhatti, Z.E., Roop, P.S., Sinha, R.: Unified functional safety assessment of
  industrial automation systems. IEEE Transactions on Industrial Informatics
  (2016)

\bibitem{bosch1999software}
Bosch, J., Molin, P.: Software architecture design: evaluation and
  transformation. In: Proceedings ECBS'99. IEEE Conference and Workshop on
  Engineering of Computer-Based Systems. IEEE (1999)

\bibitem{broy2009automotive}
Broy, M., Gleirscher, M., Kluge, P., Krenzer, W., Merenda, S., Wild, D.:
  Automotive architecture framework: Towards a holistic and standardised system
  architecture description. IEEE Computer  \textbf{42}(12) (2009)

\bibitem{bucaioni2020technical}
Bucaioni, A., Pelliccione, P.: Technical architectures for automotive systems.
  In: (ICSA). IEEE (2020)

\bibitem{cuenot2014applying}
Cuenot, P., Ainhauser, C., Adler, N., Otten, S., Meurville, F.: Applying model
  based techniques for early safety evaluation of an automotive architecture in
  compliance with the iso 26262 standard (2014)

\bibitem{Dajsuren2019}
Dajsuren, Y., van~den Brand, M. (eds.): Automotive Systems and Software
  Engineering: State of the Art and Future Trends. Springer International
  Publishing (2019)

\bibitem{dajsuren2019safety}
Dajsuren, Y., Loupias, G.: Safety analysis method for cooperative driving
  systems. In: (ICSA). IEEE (2019)

\bibitem{dajsuren2015design}
Dajsuren, Y.: On the design of an architecture framework and quality evaluation
  for automotive software systems. Ph.D. thesis, Department of Mathematics and
  Computer Science, Eindhoven University of Technology (2015)

\bibitem{davila2013report}
Davila, A.: Report on fuel consumption. SARTRE, Deliverables  (2013)

\bibitem{dobrica2002survey}
Dobrica, L., Niemela, E.: A survey on software architecture analysis methods.
  IEEE Transactions on software Engineering  \textbf{28}(7) (2002)

\bibitem{fu2018fault}
Fu, Y.: Fault injection mechanisms for validating dependability of automotive
  systems. Master's thesis, Eindhoven University of Technology (2018)

\bibitem{harrison2010pattern}
Harrison, N., Avgeriou, P.: Pattern-based architecture reviews. IEEE software
  \textbf{28}(6) (2010)

\bibitem{harrison2013using}
Harrison, N.B., Avgeriou, P.: Using pattern-based architecture reviews to
  detect quality attribute issues-an exploratory study. In: Transactions on
  Pattern Languages of Programming III. Springer (2013)

\bibitem{hasan2020fail}
Hasan, S.: Fail-Operational and Fail-Safe Vehicle Platooning in the Presence of
  Transient Communication Errors. Ph.D. thesis, M{\"a}lardalen University
  (2020)

\bibitem{hommes2012review}
Hommes, Q.V.E.: Review and assessment of the iso 26262 draft road
  vehicle-functional safety. Tech. rep., SAE Technical Paper (2012)

\bibitem{IEC61508}
IEC: {IEC} functional safety and {IEC 61508}. Standard, International
  Electrotechnical Commission (2010)

\bibitem{ISO26262-1}
ISO: {ISO 26262: 2011 - Road vehicles – Functional safety}. Standard,
  International Organization for Standardization (2011)

\bibitem{ISO26262-2}
ISO: {ISO 26262: 2018 - Road vehicles – Functional safety}. Standard,
  International Organization for Standardization (2018)

\bibitem{ISO21448}
ISO: {ISO/PAS 21448: 2019 - Road vehicles — Safety of the intended
  functionality}. Standard, International Organization for Standardization
  (2019)

\bibitem{kazman1998architecture}
Kazman, R., Klein, M., Barbacci, M., Longstaff, T., Lipson, H., Carriere, J.:
  The architecture tradeoff analysis method. In: Proceedings. Fourth IEEE
  International Conference on Engineering of Complex Computer Systems. IEEE
  (1998)

\bibitem{kochanthara21}
Kochanthara, S.: A case study on iso 26262 extension for connected driving.
  \url{https://github.com/SangeethNila/casestudy_ISO26262_extension_connected_driving}
  (2021)

\bibitem{kochanthara2020semi}
Kochanthara, S., Rood, N., Cleophas, L., Dajsuren, Y., van~den Brand, M.:
  Semi-automatic architectural suggestions for the functional safety of
  cooperative driving systems. In: (ICSA-C). IEEE (2020)

\bibitem{lee1985fault}
Lee, W.S., Grosh, D.L., Tillman, F.A., Lie, C.H.: Fault tree analysis, methods,
  and applications a review. IEEE transactions on reliability  \textbf{34}(3)
  (1985)

\bibitem{liang2015heavy}
Liang, K.Y., M{\aa}rtensson, J., Johansson, K.H.: Heavy-duty vehicle platoon
  formation for fuel efficiency. Transactions on Intelligent Transportation
  Systems  (2015)

\bibitem{mallozzi2019autonomous}
Mallozzi, P., Pelliccione, P., Knauss, A., Berger, C., Mohammadiha, N.:
  Autonomous vehicles: State of the art, future trends, and challenges. In:
  Automotive Systems and Software Engineering. Springer (2019)

\bibitem{mallozzi2016formal}
Mallozzi, P., Sciancalepore, M., Pelliccione, P.: Formal verification of the
  on-the-fly vehicle platooning protocol. In: (SERENE). Springer (2016)

\bibitem{martin2020combined}
Martin, H., Ma, Z., Schmittner, C., Winkler, B., Krammer, M., Schneider, D.,
  Amorim, T., Macher, G., Kreiner, C.: Combined automotive safety and security
  pattern engineering approach. Reliability Engineering \& System Safety
  (2020)

\bibitem{nilsson2013functional}
Nilsson, J., Bergenhem, C., Jacobson, J., Johansson, R., Vinter, J.: Functional
  safety for cooperative systems. Tech. rep., SAE Technical Paper (2013)

\bibitem{oscarsson2016applying}
Oscarsson, J., Stolz-Sundnes, M., Mohan, N., Izosimov, V.: Applying
  systems-theoretic process analysis in the context of cooperative driving. In:
  (SIES) (2016)

\bibitem{pelliccione2020beyond}
Pelliccione, P., Knauss, E., {\AA}gren, S.M., Heldal, R., Bergenhem, C., Vinel,
  A., Brunneg{\aa}rd, O.: Beyond connected cars: A systems of systems
  perspective. Science of Computer Programming  \textbf{191} (2020)

\bibitem{ploeg2014analysis}
Ploeg, J.: Analysis and design of controllers for cooperative and automated
  driving  (2014)

\bibitem{preschern2013building}
Preschern, C., Kajtazovic, N., Kreiner, C.: Building a safety architecture
  pattern system. EuroPLoP '13, ACM (2015)

\bibitem{preschern2013catalog}
Preschern, C., Kajtazovic, N., Kreiner, C., et~al.: Catalog of safety tactics
  in the light of the iec 61508 safety lifecycle. In: Proceedings of VikingPLoP
  2013 Conference (2013)

\bibitem{riel2018architectural}
Riel, A., Kreiner, C., Messnarz, R., Much, A.: An architectural approach to the
  integration of safety and security requirements in smart products and systems
  design. CIRP annals  \textbf{67}(1),  173--176 (2018)

\bibitem{roy2008methods}
Roy, B., Graham, T.N.: Methods for evaluating software architecture: A survey.
  School of Computing TR  \textbf{545} (2008)

\bibitem{saberi2018functional}
Saberi, A.K., Barbier, E., Benders, F., van~den Brand, M.: On functional safety
  methods: A system of systems approach. In: (SysCon). IEEE (2018)

\bibitem{sawade2018robust}
Sawade, O., Schulze, M., Radusch, I.: Robust communication for cooperative
  driving maneuvers. IEEE Intelligent Transportation Systems Magazine
  \textbf{10}(3),  159--169 (2018)

\bibitem{serban2018standard}
{Serban}, A.C., {Poll}, E., {Visser}, J.: A standard driven software
  architecture for fully autonomous vehicles. In: (ICSA-C) (April 2018)

\bibitem{sljivo2020guiding}
Sljivo, I., Uriagereka, G.J., Puri, S., Gallina, B.: Guiding assurance of
  architectural design patterns for critical applications. Journal of Systems
  Architecture  (2020)

\bibitem{fmea}
Stamatis, D.H.: Failure mode and effect analysis: FMEA from theory to
  execution. Quality Press (2003)

\bibitem{staron2017automotive}
Staron, M.: Automotive software architectures. Automot. Softw. Archit  (2017)

\bibitem{stoermer2003scam}
Stoermer, C., Bachmann, F., Verhoef, C.: Scam: The software architecture
  comparison analysis method. Tech. rep., Carnegie-mellon Univ Pittsburgh PA
  Software Engineering Inst (2003)

\bibitem{stoltz2019stpa}
Stoltz-Sundnes, M.: Stpa-inspired safety analysis of driver-vehicle interaction
  in cooperative driving automation (2019)

\bibitem{trego2010analysis}
Trego, T., Murray, D.: An analysis of the operational costs of trucking. In:
  Transportation Research Board 2010 Annual Meetings CD-ROM. Washington, DC.
  vol.~18, p.~20 (2010)

\bibitem{wu2004safety}
Wu, W., Kelly, T.: Safety tactics for software architecture design. In:
  (COMPSAC). IEEE (2004)

\bibitem{zalewski2017cognitive}
Zalewski, A., Borowa, K., Ratkowski, A.: On cognitive biases in architecture
  decision making. In: (ECSA). Springer (2017)

\end{thebibliography}

\end{document}